# Rayleigh surface waves of extremal elastic materials

Yu Wei[1], Yi Chen[2,3*], Wen Cheng[1], Xiaoning Liu[1] and Gengkai Hu[1*]

[1]School of Aerospace Engineering, Beijing Institute of Technology, Beijing 100081, China

[2]Institute of Nanotechnology, Karlsruhe Institute of Technology (KIT), Karlsruhe 76128, Germany.

[3]Institute of Applied Physics, Karlsruhe Institute of Technology (KIT), Karlsruhe 76128, Germany.

Corresponding author: yi.chen@partner.kit.edu (Y.C.), hugeng@bit.edu.cn (G.H.)

**Abstract:**

Extremal elastic materials here refer to a specific class of elastic materials whose elastic matrices exhibit one or more zero eigenvalues, resulting in soft deformation modes that, in principle, cost no energy. They can be approximated through artificially designed solid microstructures. Extremal elastic materials have exotic bulk wave properties unavailable with conventional solids due to the soft modes, offering unprecedented opportunities for manipulating bulk waves, e.g., acting as phonon polarizers for elastic waves or invisibility cloaks for underwater acoustic waves. Despite their potential, Rayleigh surface waves, crucially linked to bulk wave behaviors of such extremal elastic materials, have largely remained unexplored so far. In this paper, we theoretically investigate the propagation of Rayleigh waves in extremal elastic materials based on continuum theory and verify our findings with designed microstructure metamaterials based on pantographic structures. Dispersion relations and polarizations of Rayleigh waves in extremal elastic materials are derived, and the impact of higher order gradient effects is also investigated by using strain gradient theory. This study provides a continuum model for exploring surface waves in extremal elastic materials and may stimulate applications of extremal elastic materials for controlling surface waves.

# 1. Introduction

Elastic solids generally support three bulk waves or body waves, with one being longitudinally polarized and the other two transversely polarized (Graff, 2012; Achenbach, 2012). The superposition of different bulk waves can lead to surface waves that propagate along a free surface while decaying exponentially away from the surface. Rayleigh waves, first predicted by Lord Rayleigh in 1885, is a typical example (Rayleigh, 1896; Vinh and Ogden, 2004). Rayleigh waves are widely used in engineering fields, e.g., non-destructive evaluation, structural health monitoring or earthquake early warning. For nowadays very popular elastic topological insulators, their topological edge states can also be treated as Rayleigh waves as well (Pal and Ruzzene, 2017; Chen et al., 2018; Chen et al., 2019a; Zhang et al., 2019; Zhang et al., 2020; Chen et al., 2021; Zhao et al., 2020). Yet, such Rayleigh waves occur on free surfaces of artificial phononic crystals with geometry or material inhomogeneity rather than on free surface of homogenous solids, and they are robust to disorders as well as fabrication imperfections (Asbóth et al., 2016).

Rayleigh waves of isotopic solids are non-dispersive and propagate slower than both the bulk longitudinal waves and the bulk shear waves (Vinh and Ogden, 2004). The ratio between the Rayleigh wave speed and the bulk wave speed is determined by the Poisson's ratio of the solids. In fact, a Rayleigh wave can be mathematically decomposed as the summation of a bulk longitudinal wave and a bulk transversal wave (Achenbach, 2012). These two bulk waves share the same real wavenumber along the surface, while exhibit different purely imaginary wavenumbers along the surface normal direction. The free surface boundary conditions are satisfied by an appropriate relative amplitude of the two bulk waves. Due to the superposition of these two bulk waves, displacements on the free surface manifest as elliptically polarized in the plane formed by the wave vector direction and the surface normal. Interestingly, the wave's propagation direction and the displacement polarization are closely linked (Long et al., 2018; Bliokh, 2022). Material particles on the free surface rotate counter-clockwise for a right-propagating Rayleigh wave, while the rotation is reversed for a left-propagating Rayleigh wave, as required by the principle of time-reversal symmetry. This behavior has been interpreted as spin-momentum locking (Yuan et al., 2021), offering an opportunity to stimulate unidirectional surface waves by using circularly polarized excitations (Yuan et al., 2021; Zhao et al., 2022).

Rayleigh waves of ordinary elastic solids have been thoroughly investigated, while the counterpart for extremal elastic materials has largely remained unexplored. Extremal elastic materials can exhibit highly different bulk wave properties than common materials (Norris, 2008; Martin et al., 2012; Wei and Hu, 2022), and have attracted considerable attention, recently (Bückmann et al., 2014; Cai et al., 2015; Chen et al., 2016; Sun et al., 2018; Sun et al., 2019; Chen et al., 2019b; Chen and Hu, 2019). Mathematically, they are defined as elastic materials whose 6×6 elastic matrixes have $N \geq 1$ zero eigenvalues (Milton and Cherkaev, 1995). Depending on $N$, we refer them as unimode materials ($N = 1$), bimode materials ($N = 2$), trimode materials ($N = 3$), quadramode materials ($N = 4$), or pentamode materials ($N = 5$), following the nomenclature of a previous paper (Milton and Cherkaev, 1995). It's noteworthy that unimode materials, bimode materials, and quadramode materials are also referenced as monomode materials, dimode materials, or tetramode materials, respectively, in literatures following the Greek number (Groß et al., 2023a). These materials can be constructed based on squares or beams connected by extremely thin parts (Kadic et al., 2012; Kadic et al., 2014; Zheng et al., 2019; Dong et al., 2021; Groß et al., 2023b). Although the eigenvalues are typically not precisely zero, they can be engineered to be orders of magnitude smaller than others (Kadic et al., 2012). Each of the zero eigenvalue is attributed to a soft deformation mode of the material in the sense that the deformation requires conceptually zero energy. Ordinary solids correspond to $N = 0$ with no soft modes. The presence of soft modes in extremal elastic materials can drastically change their bulk wave properties. For example, pentamode materials exclusively support longitudinal waves, resembling the acoustic property of usual fluids (Norris, 2009). For quadramode materials, only one shear mode can propagate while the other shear mode with orthogonal polarization and the longitudinal mode have zero velocities (Wei et al., 2021). Extremal elastic materials have been adopted for numerous interesting applications for controlling bulk wave propagating, including phonon polarizer, acoustic cloak, and metasurface (Chen et al., 2015; Chen et al., 2017; Su et al., 2017; Groß et al., 2023a). However, investigations into Rayleigh surface waves, closely linked to bulk wave properties, in extremal elastic materials have been relatively scarce.

In this paper, we focus on the study of Rayleigh surface waves of two-dimensional (2D) extremal elastic materials, whose elasticity matrices are 3×3. In such case, we have two types of extremal elastic materials, i.e., the unimode material ($N = 1$) and the bimode material ($N = 2$). For extremal elastic materials, the usual strain energy resulting from the first order gradient of the displacement field can become very small due to the presence of soft modes (Alibert and Della Corte, 2015). However, the contribution of higher-order deformations can become dominate. Cauchy elasticity, which essentially ignores elastic energy from higher-order strain (Milton, 2002), can be insufficient to completely characterize wave property of extremal elastic materials. For instance, pantographic structures primarily derive their elastic energy from the second order gradient of the displacement field (Alibert and Della Corte, 2015). Therefore, we aim to explore the impact of higher-order gradient effects on Rayleigh waves of extremal elastic materials. Among various elastic continuum theories that accounts for high-order effects (Eringen, 1966; Milton et al., 2006; Auffray et al., 2009; Chen et al., 2014; Chen et al., 2020), we adopt strain gradient theory since it is a natural generalization of the Cauchy elastic theory through introducing the high-order displacement gradients. Moreover, Rayleigh waves in high order continuum media have also been studied, such as in micropolar or microstretch materials (Eringen, 2012), or in strain-gradient materials (Georgiadis et al., 2004; Yang et al, 2020; Zisis et al, 2023). All these studies focus on ordinary materials without easy deformation modes, however here we aim at surface waves in extremal elastic materials. Apart from the unimode materials and bimode materials, the wave properties of 2D trimode material ($N = 3$), whose Cauchy elastic matrix is entirely zero, are examined within the framework of the strain gradient theory. Finally, we design metamaterial structures exhibiting pronounced gradient effects and validate our findings based on continuum theory with the metamaterial model.

The paper is organized as follows. In section 2, we investigate surface waves on the free surface of extremal elastic metamaterial within the framework of Cauchy continuum theory. We aim to analytically derive their dispersion relations and displacement polarizations, comparing them with those of ordinary solids. In section 3, we consider the influence of higher-order deformation gradients to the propagation of surface waves for extremal elastic materials based on strain gradient theory. The impact of higher-order elastic parameters will be analyzed. Section 4 is dedicated to validating the mentioned results obtained from the continuum elasticity theory. For this purpose, a metamaterial toy model based on springs and masses is proposed to realize extremal elastic metamaterials with prominent strain gradient effects. The metamaterial is then homogenized as an effective strain-gradient medium. The surface wave properties are studied with the continuum model and are compared with direct numerical simulations with microstructures. Finally, the concluding section summarizes the findings presented in the paper.

## 2. 2D extremal materials: continuum model

The material we consider here is a homogeneous elastic material with a free surface, as shown in Fig. 1. In a 2D Cartesian coordinate system $\{x_1, x_2\}$, the equations of motion in the absence of body force and the constitutive law of a Cauchy elastic material are

$$\sigma_{ij,j} = \rho \frac{\partial^2 u_i}{\partial t^2}, \quad \sigma_{ij} = C_{ijkl} u_{k,l}, \tag{1}$$

in which, $\sigma_{ij}$ is the symmetric stress tensor, $u_i$ denotes the displacement, $t$ is time, $\rho$ and $C_{ijkl}$ represent the mass density and the fourth-order elasticity tensor, respectively. Repeated indices should be understood as Einstein summation. The comma denotes spatial derivative. The fourth-order elasticity tensor, $C_{ijkl}$, for an elastic material with $N$ zero eigenvalues can be expressed as (Milton and Cherkaev, 1995)

$$C_{ijkl} = \sum_{r=1}^{3-N} K_r S_{ij}^{(r)} S_{kl}^{(r)}, \tag{2}$$

with $K_r$ being the non-zero eigenvalues of the elastic matrix and $\mathbf{S}^{(r)}$ being a second order symmetric tensor (hard mode). In the following derivation, Eq. (2) is written as a compact form $C_{ijkl} = \sum_{r=1}^{3-N} S_{ij}^{(r)} S_{kl}^{(r)}$ for convenience, where $K_r$ is absorbed in $\mathbf{S}^{(r)}$.

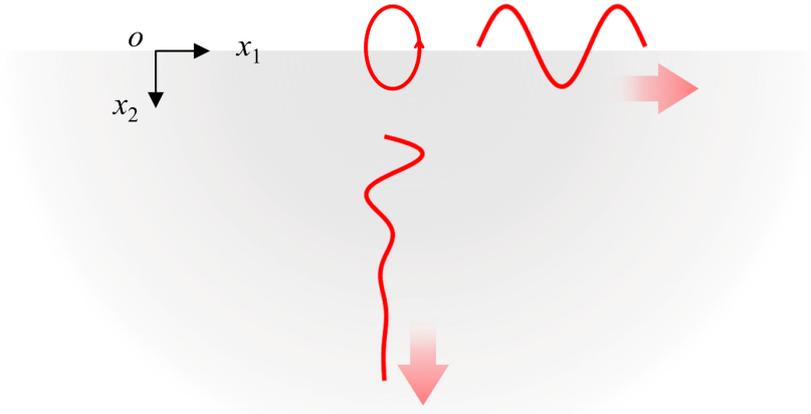

Fig. 1. Scheme of Rayleigh surface waves propagating on the free surface of a semi-infinite elastic material.

Now, we study Rayleigh surface waves propagating along a flat surface of the semi-infinite elastic material in Fig. 1. Since a Rayleigh surface mode can be regarded as the supposition of multiple evanescent modes (Achenbach, 2012), we first study possible evanescent modes in the material. We assume the following displacement form

$$\mathbf{u} = \begin{pmatrix} \hat{u}_1 \\ \hat{u}_2 \end{pmatrix} \exp(-bx_2) \exp(\mathrm{i}(kx_1 - \omega t)) = \hat{\mathbf{u}} \exp(\mathrm{i}k(\mathbf{m}\cdot\mathbf{x} - ct)), \ \hat{\mathbf{u}} = \begin{pmatrix} \hat{u}_1 \\ \hat{u}_2 \end{pmatrix}, \tag{3}$$

where $\omega$ denotes the angular frequency, $k$ the wave number, $\mathrm{i}$ the imaginary unit. For simplicity, we introduce the phase velocity $c = \omega/k$ and $\mathbf{m} = (1, +\mathrm{i}\xi)^\mathrm{T}$, with $\xi = b/k$. In the following analysis, we consider right-propagating waves, i.e., $k > 0$. Therefore, we need $\mathrm{Re}(\xi) > 0$ to ensure that the wave decays exponentially along the $+x_2$ direction. Substituting Eqs. (3) and (2) into Eq. (1) yields two homogeneous equations

$$\mathbf{Q}\cdot\hat{\mathbf{u}} = 0, \quad Q_{ik} = \sum_{r=1}^{3-N} S_{ij}^{(r)} S_{kl}^{(r)} m_l m_j - \rho c^2 \delta_{ik}, \tag{4}$$

where $\delta_{ik}$ is the Kronecker delta. To obtain a nontrivial solution, we need

$$\det(\mathbf{Q}) = 0. \tag{5}$$

Then, the solution of $\xi$ can be obtained by solving Eq. (5), and the corresponding eigenmode $\hat{\mathbf{u}}$ can be computed by substituting $\xi$ back into Eq. (4). For unimode and nullmode materials, there are generally two solutions of $\xi$ with $\mathrm{Re}\{\xi\} > 0$. The displacement for a surface wave can be written as

$$\mathbf{u} = \left(A_1 \hat{\mathbf{u}}_1 \exp(-k\xi_1 x_2) + A_2 \hat{\mathbf{u}}_2 \exp(-k\xi_2 x_2)\right) \exp(\mathrm{i}k(x_1 - ct)). \tag{6}$$

In which, $A_1$ and $A_2$ are two unknown constants. For bimode materials, there is only one solution of $\xi$ with $\mathrm{Re}\{\xi\} > 0$. This will be analyzed case by case in the following example.

Next, we will examine the traction-free conditions at the surface $x_2 = 0$, which reads

$$f_i = C_{ijkl} u_{k,l} n_j = 0, \text{ at } x_2 = 0, \tag{7}$$

where $\mathbf{n} = (0, \ 1)^{\mathrm{T}}$ is the normal vector of the free plane surface. Substituting Eq. (6) into Eq. (7) yields two homogeneous linear equations for the unknown constants $A_1$ and $A_2$

$$P_{ik} A_k = 0, \ P_{ik} = \partial f_i / \partial A_k. \tag{8}$$

Non-trivial solution for surface wave is obtained if the following condition is satisfied

$$\det(\mathbf{P}) = 0. \tag{9}$$

The Rayleigh wave velocity $c$ can be obtained by solving Eq. (9). Then, substituting $c$ back into Eq. (8) yields the constants $A_1$ and $A_2$ for the surface wave.

We note here that although Stroh formalism (Stroh, 1958) is a powerful and elegant method to study Rayleigh waves, however it cannot be applied to extremal materials, whose elasticity tensors are not positive definite. Therefore, instead of relying on Stroh formalism, we derive the Rayleigh wave using the method described above. As we pursue closed-form formula, only the case where the principal axis of the materials is parallel to $\mathbf{n}$ is considered in this paper. In such case, either the material supports Rayleigh surface waves with zero phase velocity or there is no Rayleigh surface mode. We provide detail analysis for 2D bimode extremal materials and unimode extremal materials next.

*2.1 2D bimode materials*

According to the given definition in Eq.(2), we assume the elastic tensor of a general 2D bimode material as following

$$\mathbf{C} = \mathbf{S} \otimes \mathbf{S}, \ \mathbf{S} = \begin{pmatrix} S_{11} & S_{12} \\ S_{12} & S_{22} \end{pmatrix} \neq \mathbf{0}. \tag{10}$$

Since the principal axis of the bimode material is parallel to $\mathbf{n}$, we have the following constraint

$$S_{11} S_{12} = S_{22} S_{12} = 0. \tag{11}$$

We first consider the case of $S_{12} = 0$. Assuming a displacement field of the form in Eq. (3), then substitution of the displacement field into Eq. (5) yields

$$\begin{pmatrix} S_{11}^2 - \rho c^2 & i\xi S_{11} S_{22} \\ i\xi S_{11} S_{22} & -\xi^2 S_{22}^2 - \rho c^2 \end{pmatrix} \begin{pmatrix} \hat{u}_1 \\ \hat{u}_2 \end{pmatrix} = 0. \tag{12}$$

We have following two bulk wave modes

$$\mathbf{u}_1 = \begin{pmatrix} i\xi S_{22} \\ -S_{11} \end{pmatrix} \exp(ik(x_1 + i\xi x_2)), \quad c = 0, \tag{13}$$

$$\mathbf{u}_2 = \begin{pmatrix} iS_{11} \\ -S_{22}\xi \end{pmatrix} \exp(ik(x_1 + i\xi x_2 - ct)), \quad \rho c^2 = S_{11}^2 - S_{22}^2 \xi^2. \tag{14}$$

It can be verified that the displacement field in Eq. (13) results in zero stress. Therefore, the displacement field in Eq. (13) for any Re{$\xi$} > 0 represents a surface mode but with zero phase velocity. Now, we check whether Eq. (14) can become a surface mode. Substituting the displacement field into the traction-free boundary condition yields

$$\mathbf{f} = kS_{22}(S_{11}^2 - \xi^2 S_{22}) \exp(ik(x_1 + i\xi x_2 - ct)) = 0, \quad \text{at } x_2 = 0. \tag{15}$$

We need either $S_{11}^2 - \xi^2 S_{22}^2 = 0$ or $S_{22} = 0$ for the traction-free condition. The first condition gives a displacement field as a special case of Eq. (13). The second condition leads to the following surface wave for any Re{$\xi$} > 0,

$$\mathbf{u} = \begin{pmatrix} iS_{11} \\ 0 \end{pmatrix} \exp(ik(x_1 + i\xi x_2 - ct)), \quad \rho c^2 = S_{11}^2. \tag{16}$$

This is a rather special situation as the elastic tensor only has a nonzero component $C_{1111} = S_{11}^2$ and the material behaves like a one-dimensional material. So, we discard this case.

Secondly, we consider the case of $S_{12} \neq 0$, but $S_{11} = S_{22} = 0$. Likewise, we obtain the following equation

$$\begin{pmatrix} -\xi^2 S_{12}^2 - \rho c^2 & i\xi S_{12}^2 \\ i\xi S_{12}^2 & S_{12}^2 - \rho c^2 \end{pmatrix} \begin{pmatrix} \hat{u}_1 \\ \hat{u}_2 \end{pmatrix} = 0. \tag{17}$$

We have following two bulk wave modes

$$\mathbf{u}_1 = \begin{pmatrix} i \\ \xi \end{pmatrix} \exp(ik(x_1 + i\xi x_2)), \quad c = 0, \tag{18}$$

$$\mathbf{u}_2 = \begin{pmatrix} i\xi \\ 1 \end{pmatrix} \exp(ik(x_1 + i\xi x_2 - ct)), \quad \rho c^2 = (1 - \xi^2) S_{12}^2. \tag{19}$$

It can be verified that the displacement field in Eq. (18) naturally satisfies the traction-free condition and therefore represents a surface mode with zero phase velocity for any Re{$\xi$} > 0. The second mode

Eq. (19) satisfies the traction-free condition only if $1 - \xi^2 = 0$. For this case, the displacement field is just a special example of Eq. (18).

We summarize in Table 1 the conditions that allow for surface Rayleigh waves in bimode elastic materials and the corresponding Rayleigh waves modes are given. In short, 2D bimode extremal materials with principal axis parallel to the surface only support Rayleigh surface waves with zero phase velocity. In fact, acoustic fluids can be treated as isotropic bimode materials in 2D space or isotropic pentamode materials in 3D space, which can only resist hydrostatic stress (Norris, 2008). It is known that acoustic fluids, such as water or air, cannot support propagating surface waves. This is consistent with the results in Table 1.

**Table 1** Rayleigh modes in 2D extremal materials with principal axis parallel to surface

| Type | Cases | Rayleigh modes |
|---|---|---|
| Bimode | $\mathbf{C} = \begin{pmatrix} S_{11}^2 & S_{11}S_{22} & 0 \\ S_{11}S_{22} & S_{22}^2 & 0 \\ 0 & 0 & 0 \end{pmatrix}$ | $\mathbf{u} = \begin{pmatrix} i\xi S_{22} \\ -S_{11} \end{pmatrix} \exp(ik(x_1 + i\xi x_2 - ct))$, $c = 0$, $\mathrm{Re}\{k\xi\} > 0$ |
|  | $\mathbf{C} = \begin{pmatrix} 0 & 0 & 0 \\ 0 & 0 & 0 \\ 0 & 0 & S_{12}^2 \end{pmatrix}$ | $\mathbf{u} = \begin{pmatrix} i \\ \xi \end{pmatrix} \exp(ik(x_1 + i\xi x_2 - ct))$, $c = 0$, $\mathrm{Re}\{k\xi\} > 0$ |
| Unimode | $\mathbf{C} = \begin{pmatrix} S_{11}^2 & S_{11}S_{22} & 0 \\ S_{11}S_{22} & S_{22}^2 & 0 \\ 0 & 0 & S_{12}^2 \end{pmatrix}$ $S_{11}S_{22} < 0, S_{12} \neq 0$ | $\mathbf{u} = \begin{pmatrix} i \\ \sqrt{-S_{11}/S_{22}} \end{pmatrix} \exp\left(ik\left(x_1 + i\sqrt{-S_{11}/S_{22}}\, x_2 - ct\right)\right)$, $c = 0$ |
|  | Other | No Rayleigh modes |

*2.2 2D unimode materials*

We assume the following elastic tensor for general 2D unimode materials

$$\mathbf{C} = \mathbf{S}^{(1)} \otimes \mathbf{S}^{(1)} + \mathbf{S}^{(2)} \otimes \mathbf{S}^{(2)}, \quad \mathbf{S}^{(1)} = \begin{pmatrix} S_{11}^{(1)} & S_{12}^{(1)} \\ S_{12}^{(1)} & S_{22}^{(1)} \end{pmatrix} \neq \mathbf{0}, \quad \mathbf{S}^{(2)} = \begin{pmatrix} S_{11}^{(2)} & S_{12}^{(2)} \\ S_{12}^{(2)} & S_{22}^{(2)} \end{pmatrix} \neq \mathbf{0}. \tag{20}$$

In this case, the following three constraints will hold

$$\begin{aligned} S_{11}^{(1)}S_{12}^{(1)} + S_{11}^{(2)}S_{12}^{(2)} &= 0, \\ S_{22}^{(1)}S_{12}^{(1)} + S_{22}^{(2)}S_{12}^{(2)} &= 0, \\ \mathbf{S}^{(1)} : \mathbf{S}^{(2)} = S_{11}^{(1)}S_{11}^{(2)} + S_{22}^{(1)}S_{22}^{(2)} + 2S_{12}^{(1)}S_{12}^{(2)} &= 0, \end{aligned} \tag{21}$$

in which, the first two come from the fact that principal axis of the unimode material is parallel to $\mathbf{n}$. More specifically, $C_{1112}$ and $C_{2212}$ obtained from the elastic tensor must be zero. The last one follows from the major symmetry of the elasticity tensor, which leads to orthogonal eigenvectors. It's worth noting that the elasticity tensor of unimode materials in Eq. (20) has six unknown parameters in total, while Eq. (21) provides three constraints on them. In other words, $\mathbf{S}^{(1)}$ and $\mathbf{S}^{(2)}$ are not independent. In general, two types of elasticity matrix can be obtained

$$\mathbf{C} = \begin{pmatrix} S_{11}^2 & \alpha S_{11} S_{22} & 0 \\ \alpha S_{11} S_{22} & S_{22}^2 & 0 \\ 0 & 0 & 0 \end{pmatrix}, \quad |\alpha| < 1, \; S_{11} > 0, \; S_{22} > 0. \tag{22}$$

$$\mathbf{C} = \begin{pmatrix} S_{11}^2 & S_{11} S_{22} & 0 \\ S_{11} S_{22} & S_{22}^2 & 0 \\ 0 & 0 & S_{12}^2 \end{pmatrix}, \quad S_{12} \neq 0, \; S_{11}^2 + S_{22}^2 > 0. \tag{23}$$

We first consider the elastic matrix of Eq. (22). Substituting Eq. (22) into Eq. (4) yields

$$\begin{pmatrix} S_{11}^2 - \rho c^2 & i\xi \alpha S_{11} S_{22} \\ i\xi \alpha S_{11} S_{22} & -\xi^2 S_{22}^2 - \rho c^2 \end{pmatrix} \begin{pmatrix} \hat{u}_1 \\ \hat{u}_2 \end{pmatrix} = 0. \tag{24}$$

To obtain nontrivial wave modes, the determinant of the coefficient matrix in Eq. (24) must be zero

$$S_{22}^2 \left( \alpha^2 S_{11}^2 + \rho c^2 - S_{11}^2 \right) \xi^2 + \rho c^2 \left( \rho c^2 - S_{11}^2 \right) = 0. \tag{25}$$

This equation only allows for solution of $\xi$ with $\mathrm{Re}\{\xi\} > 0$ if $(1-\alpha)S_{11}^2 < \rho c^2 < S_{11}^2$ holds. The corresponding wave mode is

$$\xi_1 = \sqrt{-\frac{\rho c^2 \left( \rho c^2 - S_{11}^2 \right)}{S_{22}^2 \left( \rho c^2 + (\alpha^2 - 1) S_{11}^2 \right)}}, \quad \mathbf{u}_1 = \begin{pmatrix} i\xi_1 \alpha S_{11} S_{22} \\ \rho c^2 - S_{11}^2 \end{pmatrix} \exp\left( ik \left( x_1 + i\xi_1 x_2 - ct \right) \right). \tag{26}$$

The traction-free condition leads to the following equation

$$S_{22} \left( S_{11} \alpha u_{11} + i S_{22} \xi_1 u_{12} \right) = 0, \quad \text{at } x_2 = 0. \tag{27}$$

In which, $u_{11}$ is the component of the vector $\mathbf{u}_1$ in Eq. (36). This equation is further simplified as

$$\left( \rho c^2 + (\alpha^2 - 1) S_{11}^2 \right) \xi_1 = 0. \tag{28}$$

Together with $\xi_1$ in Eq. (26), we obtain following solutions of $c$

$$c = 0, \quad \text{or} \quad c = \sqrt{\frac{S_{11}^2}{\rho}}, \quad \text{or} \quad c = \sqrt{(1-\alpha^2) \frac{S_{11}^2}{\rho}}. \tag{29}$$

Considering the requirement of $(1-\alpha)S_{11}^2 < \rho c^2 < S_{11}^2$, only $c = 0$ is reasonable. However, $c = 0$ results in $\xi_1 = 0$. Therefore, for the given elastic matrix in Eq. (22), there is no surface wave mode.

We then consider an elasticity matrix of the form in Eq. (23). Substituting Eq. (23) into Eq. (4) leads

$$\begin{pmatrix} S_{11}^2 - \xi^2 S_{12}^2 - \rho c^2 & i\xi\left(S_{12}^2 + S_{11}S_{22}\right) \\ i\xi\left(S_{12}^2 + S_{11}S_{22}\right) & S_{12}^2 - \xi^2 S_{22}^2 - \rho c^2 \end{pmatrix} \begin{pmatrix} \hat{u}_1 \\ \hat{u}_2 \end{pmatrix} = 0. \tag{30}$$

To obtain nontrivial bulk mode, the determinant of the coefficient matrix in Eq. (30) must be zero

$$a_0 \xi^4 + b_0 \xi^2 + c_0 = 0, \tag{31}$$

where

$$\begin{aligned} a_0 &= S_{12}^2 S_{22}^2, \\ b_0 &= 2 S_{11} S_{22} S_{12}^2 + \rho c^2 \left(S_{12}^2 + S_{22}^2\right), \\ c_0 &= S_{11}^2 S_{12}^2 - \rho c^2 \left(S_{11}^2 + S_{12}^2\right) + \rho^2 c^4. \end{aligned} \tag{32}$$

If $S_{22} = 0$, there is one solution of $\xi$ with $\text{Re}\{\xi\} > 0$ under the condition of $b_0 c_0 < 0$. From the traction-free boundary condition, we can solve $c = \sqrt{S_{11}^2/\rho}$. The mode becomes a bulk wave mode and should be discarded.

We only need to consider $S_{22} \neq 0$. Here, the solutions of Eq. (32) depend on $b_0^2 - 4a_0 c_0$. There are two pairs of degenerate solutions of $\xi$ if $b_0^2 - 4a_0 c_0 = 0$. Otherwise, there are two pairs of complex conjugated solution of $\xi$ if $b_0^2 - 4a_0 c_0 \neq 0$. We discuss the two cases separately.

For $b_0^2 - 4a_0 c_0 = 0$, we need to consider $S_{12}^2 - S_{22}^2 = 0$ or $S_{12}^2 - S_{22}^2 \neq 0$. If $S_{12}^2 - S_{22}^2 = 0$, then the condition together with $b_0^2 - 4a_0 c_0 = 0$ leads to $S_{11} + S_{22} = 0$. Then, we obtain the following wave mode from Eq. (31)

$$\begin{aligned} \xi_1 &= \sqrt{\frac{S_{22}^2 - \rho c^2}{S_{22}^2}}, \quad \mathbf{u}_1 = \begin{pmatrix} 1 \\ 0 \end{pmatrix} \exp\left(ik\left(x_1 + i\xi_1 x_2 - ct\right)\right), \\ \xi_2 &= \sqrt{\frac{S_{22}^2 - \rho c^2}{S_{22}^2}}, \quad \mathbf{u}_2 = \begin{pmatrix} 0 \\ 1 \end{pmatrix} \exp\left(ik\left(x_1 + i\xi_1 x_2 - ct\right)\right). \end{aligned} \tag{33}$$

It can be verified that, the above solution satisfies the traction-free boundary condition only if $S_{22} = 0$, which contradicts the prerequisite condition of $S_{22} \neq 0$. This means there is no surface mode when $S_{12}^2 - S_{22}^2 = 0$ and $S_{11} + S_{22} = 0$, which represents a special case of unimode material.

If $b_0^2 - 4a_0 c_0 = 0$ and $S_{12}^2 - S_{22}^2 \neq 0$, we obtain the solution of $c$

$$c = 0, \quad \text{or} \quad c = \sqrt{-4 \frac{S_{12}^2}{\rho} \frac{\left(S_{11}S_{22} + S_{22}^2\right)\left(S_{11}S_{22} + S_{12}^2\right)}{\left(S_{12}^2 - S_{22}^2\right)^2}}. \tag{34}$$

Substitution of $c = 0$ into Eq. (31) leads to the following mode

$$\xi = \sqrt{-\frac{S_{11}}{S_{22}}}, \quad \mathbf{u} = \begin{pmatrix} i\xi_1 S_{22} \\ -S_{11} \end{pmatrix} \exp\left(ik\left(x_1 + i\xi x_2 - ct\right)\right), \quad c = 0. \tag{35}$$

The above mode satisfies the traction-free boundary condition automatically, and thus represents a non-propagating Rayleigh surface modes when $S_{11}S_{22} < 0$.

For the second solution of $c$ in Eq. (34), we have the following wave mode

$$\xi_1 = \sqrt{\frac{\left(S_{12}^2 + 2S_{11}S_{22} + S_{22}^2\right)\left(2S_{12}^2 S_{22} + S_{11}S_{12}^2 + S_{11}S_{22}^2\right)}{S_{22}\left(S_{12}^2 - S_{22}^2\right)^2}},$$

$$\mathbf{u} = \begin{pmatrix} i\xi_1\left(S_{12}^2 + S_{11}S_{22}\right) \\ \xi_1^2 S_{12}^2 + \rho c^2 - S_{11}^2 \end{pmatrix} \exp\left(ik\left(x_1 + i\xi_1 x_2 - ct\right)\right).$$

(36)

From the traction-free condition, we need

$$iS_{11}u_{11} - S_{22}u_{12}\xi_1 = u_{12} + iu_{11}\xi_1 = 0, \text{ at } x_2 = 0,$$

(37)

in which, $u_{11}$ is the component of the vector $\mathbf{u}_1$ in Eq. (36). Eliminating $u_{12}$ from the two equation leads to $iu_{11}(S_{11} + S_{22}\xi_1^2) = 0$. For nontrivial solution, we require $S_{11} + S_{22}\xi_1^2 = 0$. Together with $\xi_1$ in Eq. (36), we obtain $S_{12}^2 + S_{11}S_{22} = 0$ or $S_{11} + S_{22} = 0$, with the prerequisite condition $S_{12}^2 - S_{22}^2 \neq 0$. One can verify that $S_{12}^2 + S_{11}S_{22} = 0$ leads to $\xi_1 = -\sqrt{S_{11}/S_{22}}$ and $c = 0$, which results in trivial displacement field in Eq.(37). Therefore, we only have the following nontrivial mode with $S_{11} + S_{22} = 0$, which is basically an example of Eq. (35)

$$S_{11} = -S_{22}, \quad \mathbf{u} = \begin{pmatrix} i \\ 1 \end{pmatrix} \exp\left(ik\left(x_1 + ix_2 - ct\right)\right), \quad c = 0.$$

(38)

In the above analysis, we have discussed the case of $b_0^2 - 4a_0c_0 = 0$. Now, if $a_0 > 0$ and $b_0^2 - 4a_0c_0 \neq 0$, we have four solutions of $\xi$, two with $\text{Re}\{\xi\} > 0$ and two with $\text{Re}\{\xi\} < 0$. We are only interested in the modes with $\text{Re}\{\xi\} > 0$, i.e.,

$$\xi_1 = \sqrt{\frac{-b_0 + \sqrt{b_0^2 - 4a_0c_0}}{2a_0}}, \quad \mathbf{u}_1 = \begin{pmatrix} i\xi_1\left(S_{12}^2 + S_{11}S_{22}\right) \\ \xi_1^2 S_{12}^2 + \rho c^2 - S_{11}^2 \end{pmatrix} \exp\left(ik\left(x_1 + i\xi_1 x_2 - ct\right)\right),$$

$$\xi_2 = \sqrt{\frac{-b_0 - \sqrt{b_0^2 - 4a_0c_0}}{2a_0}}, \quad \mathbf{u}_2 = \begin{pmatrix} i\xi_2\left(S_{12}^2 + S_{11}S_{22}\right) \\ \xi_2^2 S_{12}^2 + \rho c^2 - S_{11}^2 \end{pmatrix} \exp\left(ik\left(x_1 + i\xi_2 x_2 - ct\right)\right).$$

(39)

Assume a surface mode is composed of the above two mode, i.e., $\mathbf{u} = A_1\mathbf{u}_1 + A_2\mathbf{u}_2$, and substitute the displacement field into the traction-free conditions in Eqs. (6)-(9), we obtain the following equation

$$\begin{pmatrix} S_{22}\left(S_{11}u_{11} + iS_{22}\xi_1 u_{12}\right) & S_{22}\left(S_{11}u_{21} + iS_{22}\xi_2 u_{22}\right) \\ S_{12}^2\left(u_{12} + i\xi_1 u_{11}\right) & S_{12}^2\left(u_{22} + i\xi_2 u_{21}\right) \end{pmatrix} \begin{pmatrix} A_1 \\ A_2 \end{pmatrix} = 0, \text{ at } x_2 = 0,$$

(40)

in which, $u_{11}$ is component of the vector $\mathbf{u}_1$. Likewise, $u_{21}$ is component of the vector $\mathbf{u}_2$. In order to obtain a nontrivial solution of $A_1$ and $A_2$. The following condition must be satisfied

$$\left(S_{11}\hat{u}_{11} + iS_{22}\xi_1\hat{u}_{12}\right)\left(\hat{u}_{22} + i\xi_2\hat{u}_{21}\right) - \left(\hat{u}_{12} + i\xi_1\hat{u}_{11}\right)\left(S_{11}\hat{u}_{21} + iS_{22}\xi_2\hat{u}_{22}\right) = 0.$$

(41)

This equation together with the expression of $\xi_1$ and $\xi_2$ in Eq. (39) leads to an equation of $c$. The equation is too lengthy and is omitted here. Four solutions of $c$ can be obtained

$$c_1^* = 0, \quad c_2^* = \sqrt{-\frac{S_{12}^2}{\rho}\frac{4S_{22}(S_{11}+S_{22})(S_{11}S_{22}+S_{12}^2)}{(S_{12}^2-S_{22}^2)^2}}, \quad c_3^* = \sqrt{\frac{S_{11}^2}{\rho}}, \quad c_4^* = \sqrt{\frac{S_{12}^2}{\rho}\frac{S_{11}^2-S_{22}^2}{S_{12}^2-S_{22}^2}}. \tag{42}$$

Solutions $c_1^*$ and $c_2^*$ are exactly the same as in Eq. (34). Only a surface mode as in Eq. (35) with $c = 0$ is allowed when $S_{11}S_{22} < 0$. The solution $c_3^*$ corresponds to traveling longitudinal wave along the interface. $c_4^*$ is physical only if $1 < S_{11}/S_{22} < S_{12}/S_{22}$ or $0 < S_{12}/S_{22} < S_{11}/S_{22} < 1$. However, this results in pure imaginary solutions of $\xi_1$ and $\xi_2$.

Through the above analysis, we have discussed all the possibility of Rayleigh surface modes in unimode materials. The condition for Rayleigh surface modes in unimode material is summarized in Table 1. In 2D unimode extremal materials with material axis parallel to the free surface, Rayleigh modes with zero phase velocity can exist only when $S_{11}S_{22} < 0$. The polarization could be circularly polarized or elliptically polarized depending on the material parameters. All other cases do not support surface waves.

We verify the above findings by using one example of microstructure unimode material. Here, we consider a twisted kagome lattice as shown in Fig. 2. The model is composed of masses (dots) connected by linear Hooke's springs (lines). The lattice can be understood as an extremal elastic material with the effective Lamé constants $\lambda^{\text{eff}} = -\mu^{\text{eff}}$ (Hutchinson and Fleck, 2006; Sun et al., 2012; Nassar et al., 2020). According to Table 1, a surface state with zero phase velocity exists in this extreme material, $\mathbf{u} = (i, 1)^T \exp(-kx_2) \exp(ikx_1)$ for $k > 0$. Figure 2(a) shows the unit cell and a strip geometry of the twisted kagome lattice. The band structure for a strip geometry with 50 unit cells along the $\mathbf{a}_1$ direction is shown in Fig. 2(b). We obtain a flat band (blue dots) with zero frequency for the kagome lattice. The flat band is in perfect agreement with the dispersion relation of the surface state (red line) in a continuum model with homogenized material parameters. We further show in Fig. 2(c) the calculated displacement field of the surface state for the kagome model with $ka/\pi = 0.1$. The real part of the displacement vector (red arrow) and the imaginary part of the displacement vector (blue arrow) are perpendicular to each other and have nearly the same length, which indicates nearly circularly polarization.

Notice that there are three independent nodes ($n_1$, $n_2$ and $n_3$, see Fig. 2(a)) in a unit cell. We define the effective displacement of the lattice as the averaged displacement of the three nodes, which can be understood as the displacement of the geometry center of the three nodes (see Fig. 2(c)). Figure 2(d) shows the normalized displacement amplitude with respect to the depth. We obtain a good agreement between the microstructure model and the effective-medium calculation. Furthermore, in panels (e₁) and (e₂), we compare the displacement trajectory of surface particles between the homogenized model and the microstructure calculation. Again, we obtain a nice consistency between the two results. In particular, the displacement trajectory of the lattice model is nearly circular even for larger wavenumbers (see panel (e$_2$)). The calculation justifies our above findings.

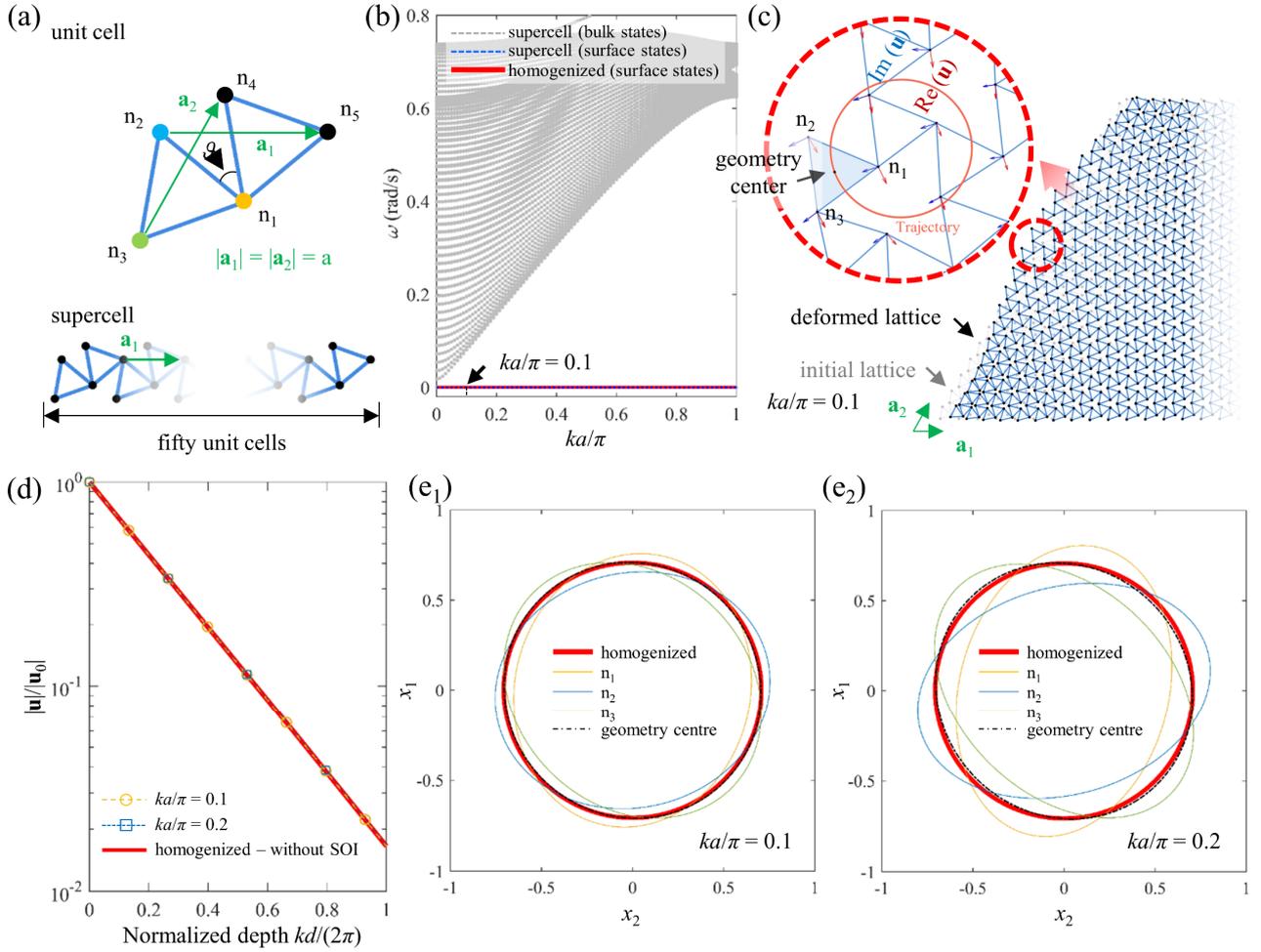

Fig. 2. Verification of circularly polarized surface waves in a twisted kagome lattice. (a) Sketches of the unit cell and the supercell; The dots represent masses and the blue lines stand for linear axial Hooke's springs. The twisted angle $\vartheta = 40°$ is chosen as an example. The lattice can be effectively treated as an unimode extremal elastic material with $\lambda^{\text{eff}} = -\mu^{\text{eff}}$. Other choices of the angle parameter lead to a similar unimode property. (b) Band structure for a strip geometry with 50 unit cells along the $\mathbf{a}_1$ direction (see panel (a)). The dashed blue line represents the surface states for microstructure model while the red line is for continuum model with homogenized material parameters. (c) Illustration of surface states with $ka/\pi = 0.1$ for the microstructure model. The red arrow and the blue arrow represent the real part and the imaginary part of the displacement, respectively. (d) Normalized displacement amplitude $|\mathbf{u}|/|\mathbf{u}_0|$ versus the normalized depth $kd/(2\pi)$ for different $ka$. $|\mathbf{u}_0|$ is the amplitude of the surface particle. (e) Comparison between trajectories of surface particles in the homogenized model and the kagome lattice with (e$_1$) $ka/\pi = 0.1$ and (e$_2$) $ka/\pi = 0.2$. Three nodes close to the free surface are chosen for the lattice (also see panel (c)). Trajectory of the geometry center of the three nodes are also plotted (dashed black line). For small wavenumber $ka/\pi = 0.1$, the three nodes for the kagome lattice exhibit nearly circular trajectories, in similar to the homogenized model. The geometry center shows a nearly ideal circular trajectory even for larger wavenumbers.

In next section, we will show that extremal materials can support propagating Rayleigh waves if second gradient effects are included.

## 3. Second gradient materials: continuum model

*3.1 Preliminary*

Second gradient elasticity is a kinetically enhanced continuum model, which can more finely characterize the influence of microstructure on the properties of materials. Different generalizations of such enhanced continuum models can be constructed (Mindlin, 1965). Here, we adopt the so-called type I formulation, in which the extra degree of freedom is defined as the second gradient of displacement (Mindlin, 1965).

In the absence of body force and body double-force, the constitutive law and the equation of motion of a homogeneous second gradient material are

$$\begin{aligned}
\left(\sigma_{ij} - M_{ijk,k}\right)_{,j} &= \rho \frac{\partial^2 u_i}{\partial t^2}, \\
\sigma_{ij} &= C_{ijkl} u_{k,l} + E_{ijklp} u_{k,lp}, \\
M_{ijk} &= E_{lpijk} u_{l,p} + D_{ijklpq} u_{l,pq},
\end{aligned} \tag{43}$$

where $M_{ijk}$ is the hyperstress tensor, $C_{ijkl}$ is the fourth-order Cauchy elasticity tensor, $D_{ijklpq}$ and $E_{ijklp}$ denote the sixth-order elasticity tensor and the fifth-order coupling elasticity tensor, respectively. They satisfy the following symmetry, $C_{ijkl} = C_{jikl} = C_{klij}$, $E_{ijklp} = E_{jiklp} = E_{ijkpl}$, and $D_{ijklpq} = D_{ikjlpq} = D_{ijklqp} = D_{lpqijk}$ (Mindlin, 1965). In 2D case, $E_{ijklp}$ vanishes for any medium with even order rotational invariance (Auffray et al., 2009). We only consider this type of medium in the paper.

Assuming the displacement field as in Eq. (3), Eq. (43) could be simplified as

$$Q_{ik}^{(S)} u_k = 0, \quad \sigma_{ij} = C_{ijkl} u_{k,l}, \quad M_{ijk} = D_{ijklpq} u_{l,pq}, \tag{44}$$

where the superscript S in parentheses represents the quantity for *second gradient elasticity*, and the generalized acoustic tensor is now written as

$$Q_{ik}^{(S)} = C_{ijkl} m_j m_l + k^2 D_{ijlkpq} m_j m_l m_p m_q - \rho c^2 \delta_{ik}. \tag{45}$$

in which, $m_i$ is the component of the vector $\mathbf{m} = (1, +\mathrm{i}\xi)^{\mathrm{T}}$. Then, the two bulk wave phase velocities and polarization can be obtained by solving the eigenvalue problem of Eqs. (44) and (45).

Naturally, the method of analyzing Rayleigh waves in the Section 2 for Cauchy elasticity can be extended to the second gradient elasticity. A non-trivial solution exists if

$$\det\left(\mathbf{Q}^{(S)}\right) = 0. \tag{46}$$

If the principal axis of the material is parallel to the surface **n**, there are usually four pairs of solutions, which can be further divided into two groups based on the sign of Re{ξ}. The general displacement solution for a free surface wave has the following form

$$\mathbf{u} = \sum_{i=1}^{4} A_i \hat{\mathbf{u}}_i \exp\left(ik\left(x_1 + i\xi_i x_2 - ct\right)\right). \tag{47}$$

For deriving surface Rayleigh waves, the following boundary conditions at the free surface $x_2 = 0$ (Mindlin, 1964) must be satisfied

$$f_i^{(S)} = \left(C_{ijkl} u_{k,l} - \left(2\delta_{pr} - n_p n_r\right) D_{ijpklq} u_{k,lqr}\right) n_j = 0,$$
$$R_i^{(S)} = D_{ijpklq} n_j n_p u_{k,lq} = 0. \tag{48}$$

Then, substituting Eq. (47) into Eq. (48) yields four homogeneous equations for the unknown constants $A_1$, $A_2$, $A_3$ and $A_4$

$$\begin{pmatrix} f_i^{(S)} \\ R_i^{(S)} \end{pmatrix} = P_{ik}^{(S)} A_k = 0, \quad P_{ik}^{(S)} = \begin{pmatrix} \partial f_i^{(S)} / \partial A_k \\ \partial R_i^{(S)} / \partial A_k \end{pmatrix}. \tag{49}$$

A non-trivial free surface wave requires

$$\det\left(\mathbf{P}^{(S)}\right) = 0. \tag{50}$$

The Rayleigh wave velocity and the displacement fields for a semi-infinite second gradient elasticity material can be obtained by solving Eq. (50).

## 3.2 Isotropic Cauchy continuum with second gradient effect

We consider a general isotropic Cauchy continuum with isotropic second gradient elasticity. The fourth-order Cauchy elastic tensor can be expressed as

$$C_{ijkl} = \lambda \delta_{ij}\delta_{kl} + \mu\left(\delta_{ik}\delta_{jl} + \delta_{il}\delta_{jk}\right), \tag{51}$$

where $\lambda$ and $\mu$ are the Lamé constants. For the sixth-order elasticity tensor (Auffray et al. (2015)), there are four independent parameters ($g_1$, $g_2$, $g_3$ and $g_4$)

$$\begin{aligned} D_{ijklpq} &= g_1\left(\delta_{ij}\delta_{kl}\delta_{pq} + \delta_{ik}\delta_{jl}\delta_{pq} + \delta_{ip}\delta_{jk}\delta_{lq} + \delta_{iq}\delta_{jk}\delta_{lp}\right) \\ &+ g_2\left(\delta_{ij}\delta_{kp}\delta_{lq} + \delta_{ij}\delta_{kq}\delta_{lp} + \delta_{ik}\delta_{jp}\delta_{lq} + \delta_{ik}\delta_{jq}\delta_{lp}\right) \\ &+ g_3\delta_{il}\delta_{jk}\delta_{pq} + g_4\left(\delta_{il}\delta_{jp}\delta_{kq} + \delta_{il}\delta_{jq}\delta_{kp}\right). \end{aligned} \tag{52}$$

Substituting Eqs. (51) and (52) into Eq. (46) yields the solution of $\xi$ and the corresponding mode $\hat{\mathbf{u}}$

$$\xi_1 = \sqrt{1 + \frac{1}{2}\frac{c_L^2}{c_{LS}^2} - \sqrt{\frac{1}{4}\frac{c_L^4}{c_{LS}^4} + \frac{c_R^2}{c_{LS}^2}}}, \hat{\mathbf{u}}_1 = \begin{pmatrix} 1 \\ i\xi_1 \end{pmatrix}, \quad \xi_2 = \sqrt{1 + \frac{1}{2}\frac{c_T^2}{c_{TS}^2} - \sqrt{\frac{1}{4}\frac{c_T^4}{c_{TS}^4} + \frac{c_R^2}{c_{TS}^2}}}, \hat{\mathbf{u}}_2 = \begin{pmatrix} \xi_2 \\ i \end{pmatrix},$$

$$\xi_3 = \sqrt{1 + \frac{1}{2}\frac{c_L^2}{c_{LS}^2} + \sqrt{\frac{1}{4}\frac{c_L^4}{c_{LS}^4} + \frac{c_R^2}{c_{LS}^2}}}, \hat{\mathbf{u}}_3 = \begin{pmatrix} 1 \\ i\xi_3 \end{pmatrix}, \quad \xi_4 = \sqrt{1 + \frac{1}{2}\frac{c_T^2}{c_{TS}^2} + \sqrt{\frac{1}{4}\frac{c_T^4}{c_{TS}^4} + \frac{c_R^2}{c_{TS}^2}}}, \hat{\mathbf{u}}_2 = \begin{pmatrix} \xi_4 \\ i \end{pmatrix},$$

(53)

In which,

$$c_L = \sqrt{\frac{\lambda + 2\mu}{\rho}}, \quad c_T = \sqrt{\frac{\mu}{\rho}}, \quad c_{LS} = k\sqrt{\frac{4g_1 + 4g_2 + g_3 + 2g_4}{\rho}}, \quad c_{TS} = k\sqrt{\frac{g_3 + 2g_4}{\rho}}. \tag{54}$$

Then, we have the matrix $\mathbf{P}$ with following components

$$P_{11}^{(S)} = (4g_1 + 4g_2 + g_3 + 2g_4)\xi_1^3 - (4g_1 + 4g_2 + g_3 + 4g_4 + 2\mu/k^2)\xi_1,$$

$$P_{21}^{(S)} = (4g_1 + 4g_2 + g_3 + 2g_4)\xi_1^4 + 2g_1 + 4g_2 + \lambda/k^2$$
$$- (6g_1 + 8g_2 + g_3 + 4g_4 + (\lambda + 2\mu)/k^2)\xi_1^2$$

$$P_{31}^{(S)} = (2g_1 + g_3 + 2g_4)\xi_1^2 - 2g_1 - g_3,$$

$$P_{41}^{(S)} = (4g_1 + 4g_2 + g_3 + 2g_4)\xi_1^3 - (4g_1 + 4g_2 + g_3)\xi_1,$$

$$P_{12}^{(S)} = ((g_3 + 2g_4)\xi_2^4 + (2g_1 - g_3 - 4g_4)\xi_2^2 - (\xi_2^2 + 1)\mu/k^2 - 2g_1)/\xi_2,$$

$$P_{22}^{(S)} = (g_3 + 2g_4)\xi_2^2 - g_3 - 4g_4 - (2\mu)/k^2,$$

$$P_{32}^{(S)} = (g_3 + 2g_4)\xi_2^2 - g_3,$$

$$P_{42}^{(S)} = ((2g_1 + g_3 + 2g_4)\xi_2^2 - 2g_1 - g_3)/\xi_2,$$

$$P_{13}^{(S)} = (4g_1 + 4g_2 + g_3 + 2g_4)\xi_3^3 - (4g_1 + 4g_2 + g_3 + 4g_4 + 2\mu/k^2)\xi_3,$$

$$P_{23}^{(S)} = (4g_1 + 4g_2 + g_3 + 2g_4)\xi_3^4 + 2g_1 + 4g_2 + \lambda/k^2$$
$$- (6g_1 + 8g_2 + g_3 + 4g_4 + (\lambda + 2\mu)/k^2)\xi_3^2$$

$$P_{33}^{(S)} = (2g_1 + g_3 + 2g_4)\xi_3^2 - 2g_1 - g_3,$$

$$P_{43}^{(S)} = (4g_1 + 4g_2 + g_3 + 2g_4)\xi_3^3 - (4g_1 + 4g_2 + g_3)\xi_3,$$

$$P_{14}^{(S)} = ((g_3 + 2g_4)\xi_4^4 + (2g_1 - g_3 - 4g_4)\xi_4^2 - (\xi_4^2 + 1)\mu/k^2 - 2g_1)/\xi_4,$$

(55)

$$P_{24}^{(S)} = (g_3 + 2g_4)\xi_4^2 - g_3 - 4g_4 - (2\mu)/k^2,$$

$$P_{34}^{(S)} = (g_3 + 2g_4)\xi_4^2 - g_3,$$

$$P_{44}^{(S)} = ((2g_1 + g_3 + 2g_4)\xi_4^2 - 2g_1 - g_3)/\xi_4.$$

The Rayleigh wave velocity as well as the displacement fields can be obtained by solving $\det(\mathbf{P}^{(S)}) = 0$. Closed formula cannot be obtained and numerical algorithm is needed to derive the solution.

We show two examples of extremal elastic materials with second gradient effect and analyze their surface wave property. The first example is

$$\lambda = 100, \; \mu = 0, \; \{g_1, g_2, g_3, g_4\} = \{0.0674, \; 0, \; 1.7977, \; 1.9326\}, \tag{56}$$

i.e., a bimode extremal material with second gradient effect. The second gradient parameters are chosen based on the proposed microstructure in Section 4 with the parameters $\theta = 44°$ and $M_{rs} = 1.0768$. The choice of $\mu = 0$ corresponds to incompressible materials. In this case, as shown in Fig. 3(a)-(b), Rayleigh waves propagate at a velocity close to that of transverse wave and increase with the increase of the wave number. Besides, the ellipticity of the polarization trajectories of the surface particles remains essentially small.

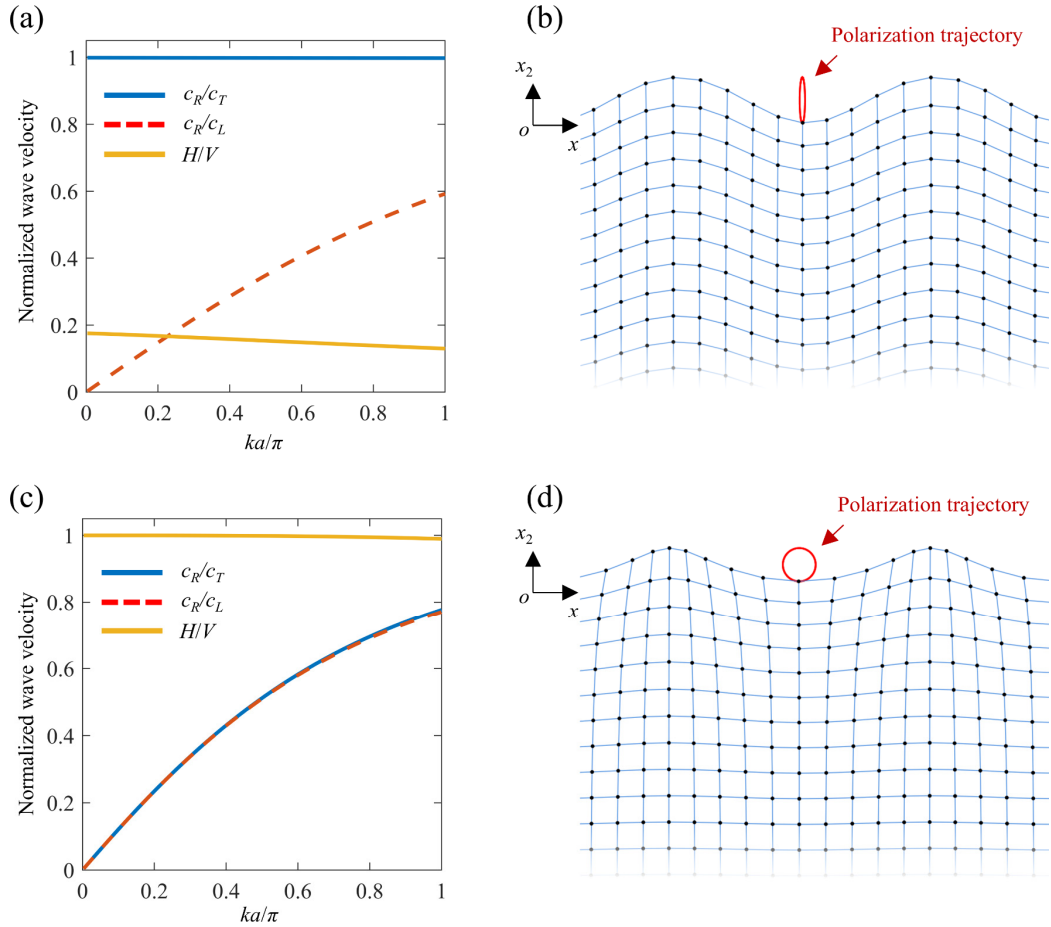

Fig. 3. The normalized Rayleigh wave velocity and the ellipticity ($H/V$) of surface particle polarization as function of the normalized wave number for a bimode extremal material with isotropic second gradient effect, $\{\lambda, \mu, g_1, g_2, g_3, g_4\}$ = $\{100, 0, 0.0674, 0, 1.7977, 1.9326\}$. (b) Illustration of Rayleigh surface wave with $ka/\pi = 0.2$. (c)-(d): Same as (a)-(b) but for an unimode extremal material with isotropic second gradient effect $\{\lambda, \mu, g_1, g_2, g_3, g_4\}$ = $\{-100, 100, 0.0674, 0, 1.7977, 1.9326\}$.

The second example is

$$\lambda = -100, \; \mu = 100, \; \{g_1, g_2, g_3, g_4\} = \{0.0674, \; 0, \; 1.7977, \; 1.9326\}, \tag{57}$$

i.e., a unimode extremal material with second gradient effect. The second gradient parameters are the same as in previous sample. The choice of $\lambda < 0$ corresponds to an auxetic material. As shown in Fig. 3(c)-(d), the normalized Rayleigh wave velocity gradually increases from zero with the increase of the wave number. In similar to the bimode material in Section 2, the polarization of surface particles exhibits nearly circular behavior.

*3.3 Isotropic pure second gradient continuum*

In the following, we consider isotropic pure second gradient continuum, i.e., the lame constant $\lambda$ and $\mu$ are both zeros. More specifically, we focus on the case of $g_2 = 0$, which can be realized by our proposed microstructure model in next section. In general, Rayleigh waves with different polarizations can be realized by appropriate choice for the four independent material constants.

We consider two examples with drastically different Rayleigh wave properties. First, $g_2 = g_3 = 0$ and $g_1/g_4 = 0.5$, the following surface mode is obtained

$$\hat{\mathbf{u}} = (0.2503, \ -0.9682i)^\mathrm{T}, \ \text{at} \ x_2 = 0, \ c_R/c_{TS} \approx 0.9852. \tag{58}$$

which indicates that the particles at the surface are elliptically polarized. This will be further demonstrated in the Section 4.2 using microstructure.

Another example is $g_1 = g_2 = 0$, $g_3/g_4 = 1$. In this case, analytical solution of the Rayleigh mode can be derived. The transverse and longitudinal waves are degenerate, i.e., $c_{TS} = c_{LS}$, as given by Eq. (53). Two degenerate Rayleigh surface modes can be obtained

$$\mathbf{u} = \begin{pmatrix} 1 \\ 0 \end{pmatrix} \left[ \exp\left(-k\sqrt{1 - \frac{2\sqrt[4]{5}}{3}} x_2\right) - \frac{1-\sqrt[4]{5}}{1+\sqrt[4]{5}} \exp\left(-k\sqrt{1 + \frac{2\sqrt[4]{5}}{3}} x_2\right) \right] \exp(ik(x_1 - c_R t)),$$

$$\mathbf{u} = \begin{pmatrix} 0 \\ 1 \end{pmatrix} \left[ \exp\left(-k\sqrt{1 - \frac{2\sqrt[4]{5}}{3}} x_2\right) - \frac{1-\sqrt[4]{5}}{1+\sqrt[4]{5}} \exp\left(-k\sqrt{1 + \frac{2\sqrt[4]{5}}{3}} x_2\right) \right] \exp(ik(x_1 - c_R t)), \tag{59}$$

$$c_R = \frac{\omega}{k} = 2\sqrt[4]{5}\sqrt{\frac{g_4}{3\rho}} k,$$

which indicate there exist two propagating Rayleigh waves. This is rather different from one Rayleigh surface waves in Cauchy elastic material. The two surface waves are both linearly polarized, one with displacement vector parallel to the free surface and the other perpendicular to the free surface. This will be evidenced in later calculations with proposed microstructure model.

Four general parameters of $g_1$, $g_2$, $g_3$ and $g_4$, we summarize in Figs. 4(a)-(c) the numerically obtained Rayleigh wave velocity and the so-called ellipticity (Malischewsky and Scherbaum, 2004), respectively. The ellipticity is defined as the ratio between the surface particle's displacement component ($H$) along the surface and the displacement component ($V$) perpendicular to the surface. It can be seen that with the decrease of $g_1/g_4$ and the increase of $g_3/g_4$, the normalized Rayleigh wave velocity increases and the ellipticity decreases. As shown in Eq. (54), when $g_2 = 0$ and $g_1/g_4$ decreases, the difference between the transverse and longitudinal wave velocity gradually decreases. In other words, a medium with comparable transverse and longitudinal wave velocities allows larger normalized Rayleigh wave velocity and smaller ellipticity. Figure 4(d) also illustrate the comparison between the continuum theory and a microstructure model, which will be detailed in the next section. The red dashed line in Fig. 4(a)-(c) corresponds to available parameters of our proposed microstructure model, which will be detailed in Section 4.

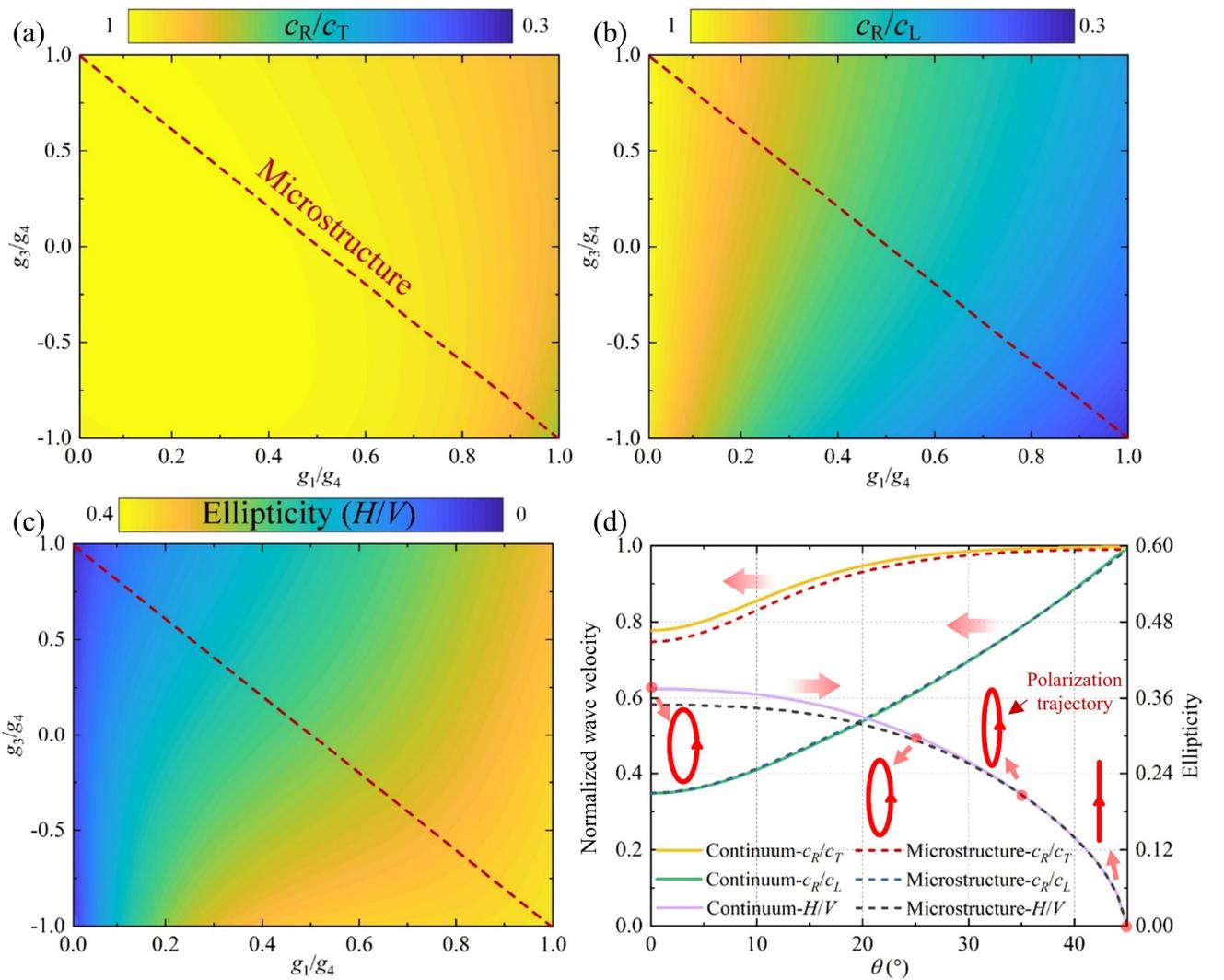

Fig. 4. (a)-(c) The normalized Rayleigh wave velocity and the ellipticity ratio ($H/V$) of surface particle polarization of pure isotropic second gradient elasticity continuum media. (d) Comparison between the continuum model and the microstructure model proposed in Section 4 with the effective parameters along the red dashed lines in (a) - (c). Here, all the results correspond to the long-wavelength limit.

We remark that in the above second gradient model, a second-order inertia (SOI) term is omitted as we are concerned with the low frequency range. While for higher frequencies, the second-order inertia tensor $J_{ijkl}$ for isotropic media (Rosi et al., 2018; Rosi and Auffray, 2019) should be included

$$J_{ijkl} = \kappa \delta_{ij}\delta_{kl} + \eta \left( \delta_{ik}\delta_{jl} + \delta_{il}\delta_{jk} \right), \tag{60}$$

in which, $\kappa$ and $\eta$ are two generalized "Lamé" parameters. The corresponding generalized acoustic tensor and the surface traction become

$$\begin{aligned} Q_{ik}^{(S)} &= \left( C_{ijkl} - \omega^2 J_{ijkl} \right) m_j m_l + k^2 D_{iqpklj} m_j m_l m_p m_q - \rho c^2 \delta_{ik}. \\ f_i^{(S)} &= \left[ \left( C_{ijkl} - \omega^2 J_{ijkl} \right) u_{k,l} - \left( 2\delta_{pr} - n_p n_r \right) D_{ijpklq} u_{k,lqr} \right] n_j = 0. \end{aligned} \tag{61}$$

In our following results, we also compare the results without the second-order inertia term and with the second-order inertia term. In general, discrepancy between the two cases is negligible at low frequency range, while the one with second-order inertia term is closer to microstructure calculations for higher frequencies.

## 4. Second gradient materials: microstructure model

### 4.1 Isotropic second gradient elasticity model

We employ the pantographic microstructures (Seppecher et al., 2011; Alibert and Della Corte, 2015) to build a microstructure model of 2D isotropic pure second gradient elastic material. The pantographic structure is consisted of rigid rods (thick blue and red lines) connected by torsion springs and ideal joints, as shown in Fig. 5(a). Black dots and hollow gay dots represent explicit degrees of freedom (DOFs) and hidden DOFs, respectively. The torsion springs with spring constant $M_{rs}$ and relaxation angle $\alpha_{rs} = \pi$ serve to couple rods of the same color.

Now, we construct an isotropic pure second gradient model based on the 1D pantographic strip in Fig. 5(a). We first define two group of substructures (Fig. 5(b)), substructure group 1 with lattice vector $\{\mathbf{a}_1^{(1)}, \mathbf{a}_2^{(1)}\}$ and substructure group 2 with lattice vector $\{\mathbf{a}_1^{(2)}, \mathbf{a}_2^{(2)}\}$, respectively. Substructure group 1 contains 1D pantographic strips with an angle of $\pi/3$ between them. Their explicit nodes form a honeycomb lattice. The substructure group 2 is obtained by rotating the substructure group 1 counterclockwise by $\pi/6$ and then scaled up by a factor of $2/\sqrt{3}$. A unit cell of the substructure group 1 and the substructure group 1 is shown in the light blue and red hexagon, respectively. Notice that the stiffness of the torsion springs is not scaled. Consequently, a relationship between the translational bases of the two groups is established

$$\mathbf{a}_1^{(2)} = \frac{2}{3}\left( \mathbf{a}_1^{(1)} + \mathbf{a}_2^{(1)} \right), \quad \mathbf{a}_2^{(2)} = \frac{2}{3}\left( 2\mathbf{a}_2^{(1)} - \mathbf{a}_1^{(1)} \right). \tag{62}$$

For convenience, hexagons (areas occupied by a unit cell) and solid dots (explicit nodes) are used to represent the two groups. We obtain a composite model, with lattice vectors $\mathbf{a}_1 = 2\mathbf{a}_1^{(1)}$ and $\mathbf{a}_2 = 2\mathbf{a}_2^{(1)}$ by combining the two groups of substructures. The two substructures are bonded together at their common explicit nodes. The composite lattice represented by explicit nodes alone is shown in the right of Fig. 5(b). For clarity, we show in Fig. 5(c) a different representation of the unit cell for the isotropic pure second gradient model, which is consisted of 4 unit cells from group 1 and three unit cells from group 2. A 3D model of the unit cell is rendered in Appendix A. A unit cell has in total 6 explicit nodes, four of which are from the group 1 (black dots) and the rest from the group 2 (red dots). Similarly, a unit cell contains 42 implicit nodes.

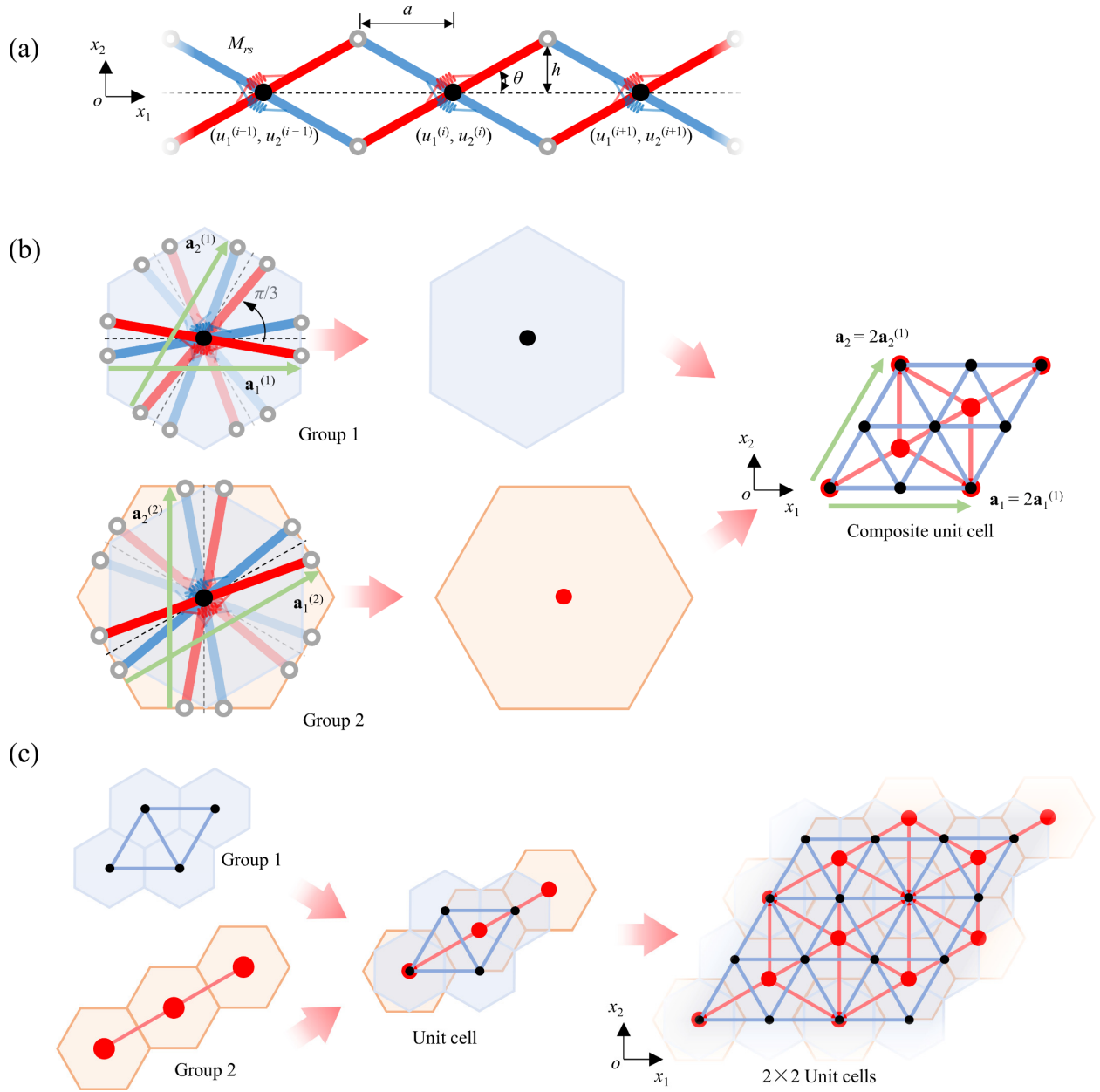

Fig. 5. An isotropic second gradient model with pantographic structures. (a) 1D pantographic structure consisted of rigid rods (thick blue and red lines) coupled by torsion springs, with spring constant $M_{rs}$, and ideal joints (hollow gray dots). Elastic energy of the system can be written in terms of the displacement $(u_1^{(i)}, u_2^{(i)})$ of the explicit nodes (black dots).

Geometric parameters are as indicated. (b) Illustration of structure composition for a 2D isotropic second gradient model. The substructure group 1 is obtained by arrange the 1D pantographic structure in a honeycomb lattice. The structure shown in the blue hexagon represents a unit cell of the substructure group 1. The substructure group 2 is similar to group 1 but rotated by $\pi/6$ and scaled up by a factor of $2/\sqrt{3}$. A composite lattice is obtained combing the two groups of substructures at their common explicit nodes (black and red dots). A unit cell of the composite lattice is schematically represented as the figure on the right. (c) A different choice of unit cell for the composite lattice. The unit cell is composed from 4 unit cells from substructure group 1 and 3 unit cells from substructure group 2. A 2 by 2 array of the composite unit cell is also shown.

We derive the effective strain gradient parameters of the microstructure model based on strain energy density. We first analyze the elastic energy in the 1D pantographic strip geometry (see Fig. 5(a)). The displacement of the explicated nodes is denoted as $\mathbf{u}^{(i)} = (u_1^{(i)}, u_2^{(i)})$. The strain energy stored in the two torsion springs at node $(i)$ can be expressed as

$$E = \frac{M_{rs}}{4} \frac{1}{h^2} \left( \left( \left( \mathbf{u}^{(i-1)} + \mathbf{u}^{(i+1)} - 2\mathbf{u}^{(i)} \right) \cdot \mathbf{e}_1 \right)^2 + \frac{h^2}{a^2} \left( \left( \mathbf{u}^{(i-1)} + \mathbf{u}^{(i+1)} - 2\mathbf{u}^{(i)} \right) \cdot \mathbf{e}_2 \right)^2 \right), \tag{63}$$

in which, $\mathbf{e}_1$ and $\mathbf{e}_2$ are unit vectors along the two axes of the coordinate system, respectively. The substructure group 1 (see Fig. 5(b)) is constructed from the 1D pantographic strip geometry. Their explicit nodes form a triangular lattice with the lattice vectors $\mathbf{a}_1^{(1)}$ and $\mathbf{a}_2^{(1)}$. We denote the explicit nodes by two numbers $\{i, j\}$. Now, the elastic energy in the unit cell at the node $\{i, j\}$ is stored in torsion springs that connects those rods between nodes $\{i, j\}$ and its six nearest-neighboring nodes. Therefore, we can write the strain energy density as

$$W_1 = \frac{M_{rs}}{4h^2 A_{\text{cell}}} \left( \begin{array}{l} \left( \left( \mathbf{u}^{(i-1,j)} + \mathbf{u}^{(i+1,j)} - 2\mathbf{u}^{(i,j)} \right) \cdot \mathbf{e}_1 \right)^2 + \frac{h^2}{a^2} \left( \left( \mathbf{u}^{(i-1,j)} + \mathbf{u}^{(i+1,j)} - 2\mathbf{u}^{(i,j)} \right) \cdot \mathbf{e}_2 \right)^2 \\ + \left( \left( \mathbf{u}^{(i,j-1)} + \mathbf{u}^{(i,j+1)} - 2\mathbf{u}^{(i,j)} \right) \cdot \mathbf{e}_3 \right)^2 + \frac{h^2}{a^2} \left( \left( \mathbf{u}^{(i,j-1)} + \mathbf{u}^{(i,j+1)} - 2\mathbf{u}^{(i,j)} \right) \cdot \mathbf{e}_4 \right) \\ + \left( \left( \mathbf{u}^{(i+1,j-1)} + \mathbf{u}^{(i-1,j+1)} - 2\mathbf{u}^{(i,j)} \right) \cdot \mathbf{e}_5 \right)^2 + \frac{h^2}{a^2} \left( \left( \mathbf{u}^{(i+1,j-1)} + \mathbf{u}^{(i-1,j+1)} - 2\mathbf{u}^{(i,j)} \right) \cdot \mathbf{e}_6 \right) \end{array} \right), \tag{64}$$

where, $\mathbf{e}_3 = \{\cos(\pi/3), \sin(\pi/3)\}$, $\mathbf{e}_4 = \{\cos(-\pi/6), \sin(-\pi/6)\}$, $\mathbf{e}_5 = \{\cos(2\pi/3), \sin(2\pi/3)\}$, $\mathbf{e}_6 = \{\cos(\pi/6), \sin(\pi/6)\}$, and $A_{\text{cell}} = 2\sqrt{3}a^2$. We consider the long-wavelength limit condition, then the displacement field can be approximated by Taylor series (Chen et al., 2014)

$$\mathbf{u}^{(i+m,j+n)} = \mathbf{u}^{(i,j)} + (\mathbf{u}\nabla) \cdot d\mathbf{x} + \frac{1}{2!}(\mathbf{u}\nabla\nabla) \cdot (d\mathbf{x})^2 + o((d\mathbf{x})^2), \quad d\mathbf{x} = m\mathbf{a}_1^{(1)} + n\mathbf{a}_2^{(1)}. \tag{65}$$

Substitution of the above expansion into Eq. (64) gives the strain energy density in the substructure group 1

$$W_1 = \frac{\sqrt{3}M_{rs}}{8} \begin{pmatrix} \frac{a^2}{h^2}\left(4u_{1,11}^2 + 4u_{2,22}^2 + \left(2u_{1,12} + u_{2,11} + u_{2,22}\right)^2 + \left(2u_{2,12} + u_{1,11} + u_{1,22}\right)^2\right) \\ + 4u_{1,22}^2 + 4u_{2,11}^2 + \left(u_{1,11} + u_{1,22} - 2u_{2,12}\right)^2 + \left(u_{2,11} + u_{2,22} - 2u_{1,12}\right)^2 \end{pmatrix}. \qquad (66)$$

Likewise, strain energy density $W_2$ for the substructure group 2 can be obtained. Here, we assume the same displacement field expansion as in Eq. (65) for explicit nodes of the substructure group 2, though only part of explicit nodes in substructure group 2 overlap with those of substructure group 1. The total energy density is the summation of the two $W_{\text{tot}} = W_1 + W_2$. The effective material parameters are obtained by comparing the energy density with that of strain gradient elasticity

$$W_{\text{tot}} = \frac{1}{2}C_{ijkl}u_{i,j}u_{k,l} + E_{ijklp}u_{i,j}u_{k,lp} + \frac{1}{2}D_{ijklpq}u_{i,jk}u_{l,pq}. \qquad (67)$$

The strain energy density results in $C_{ijkl} = 0$, $E_{ijklp} = 0$ and an isotropic six order elasticity tensor $D_{ijklpq}$ with the following parameters

$$g_1 = \frac{\sqrt{3}M_{rs}}{2}\left(\frac{1}{\tan^2\theta} - 1\right),\ g_2 = 0,\ g_3 = \frac{\sqrt{3}M_{rs}}{2}\left(3 - \frac{1}{\tan^2\theta}\right),\ g_4 = \frac{\sqrt{3}M_{rs}}{2}\left(1 + \frac{1}{\tan^2\theta}\right). \qquad (68)$$

We obtain the effective density $\rho^{\text{eff}}$ from mass average, $\rho^{\text{eff}} = 6m_0/A_{\text{cell}} = \sqrt{3}m_0/(4a^2)$, where $m_0$ is the mass of each explicit node, since we consider a frequency range far below the resonance frequency.

*4.2 Comparison between discrete model and continuum model*

We first study bulk wave properties of the proposed microstructure material. We compare in Fig. 6(a₁) the obtained band structures from the microstructure model (blue and green dots), with angle $\theta = \pi/6$, and the strain-gradient continuum model (red and black lines). The generalized "Lamé" parameters accounting for the second-order inertia terms are obtained by fitting the dispersion relations (Boutin and Auriault, 1993; Rosi et al., 2018; Rosi and Auffray, 2019). As can be seen, at low frequency range, or with wavenumbers up to $0.4\pi/a$, the second-order inertia term is negligible. For larger wavenumbers, even approaching the edge of the Brillouin zone boundary, the second-order inertia offers a nice agreement to the microstructure (compare black lines and dots). As in Cauchy media, the displacement vectors for the longitudinal waves and the transverse waves are parallel and perpendicular to the wave vector, respectively (see Fig. 6(b₁)). The results for the angle parameter $\theta = \pi/4$ are shown in Figs. 6(a₂) and 6(b₂). Here, the agreement between the continuum model and the microstructure calculation is again excellent.

In the following, we study the surface wave property in the two strain-gradient materials. The same second-order inertia parameters are used. Firstly, we consider the case of $\theta = \pi/6$. The resulting homogenized parameters $g_2 = g_3 = 0$, $g_1 = 0.5g_4 = 0.005\sqrt{3}$ corresponds to the first case in Section 3.3. Therefore, this isotropic pantographic structure could have an elliptically polarized Rayleigh wave, as suggested in the form of Eq. (58) and Fig. 4(d). In Fig. 7(a), we compare the calculated band structures (See Appendix B for method) for a strip geometry by using microstructures and the strain-gradient model with the above effective material parameters. For the surface state, we obtain a perfect agreement between the continuum model (red line) and the microstructure calculation (blue line) for wavenumbers up to $0.4\pi/a$. By including the second-order inertia term in the continuum model, the discrepancy between the continuum model (black line) and the microstructure calculation is further decreased. Furthermore, the displacement field and the particle trajectory of the surface states obtained from the three calculations are in excellent agreement with each other (see panels (b), (d)) too. As predicted in previous section, here we obtain elliptically polarized displacement trajectory, with the longer axis perpendicular to the surface. The obtained displacement decay profiles (see panel (c)) are also in good consistency.

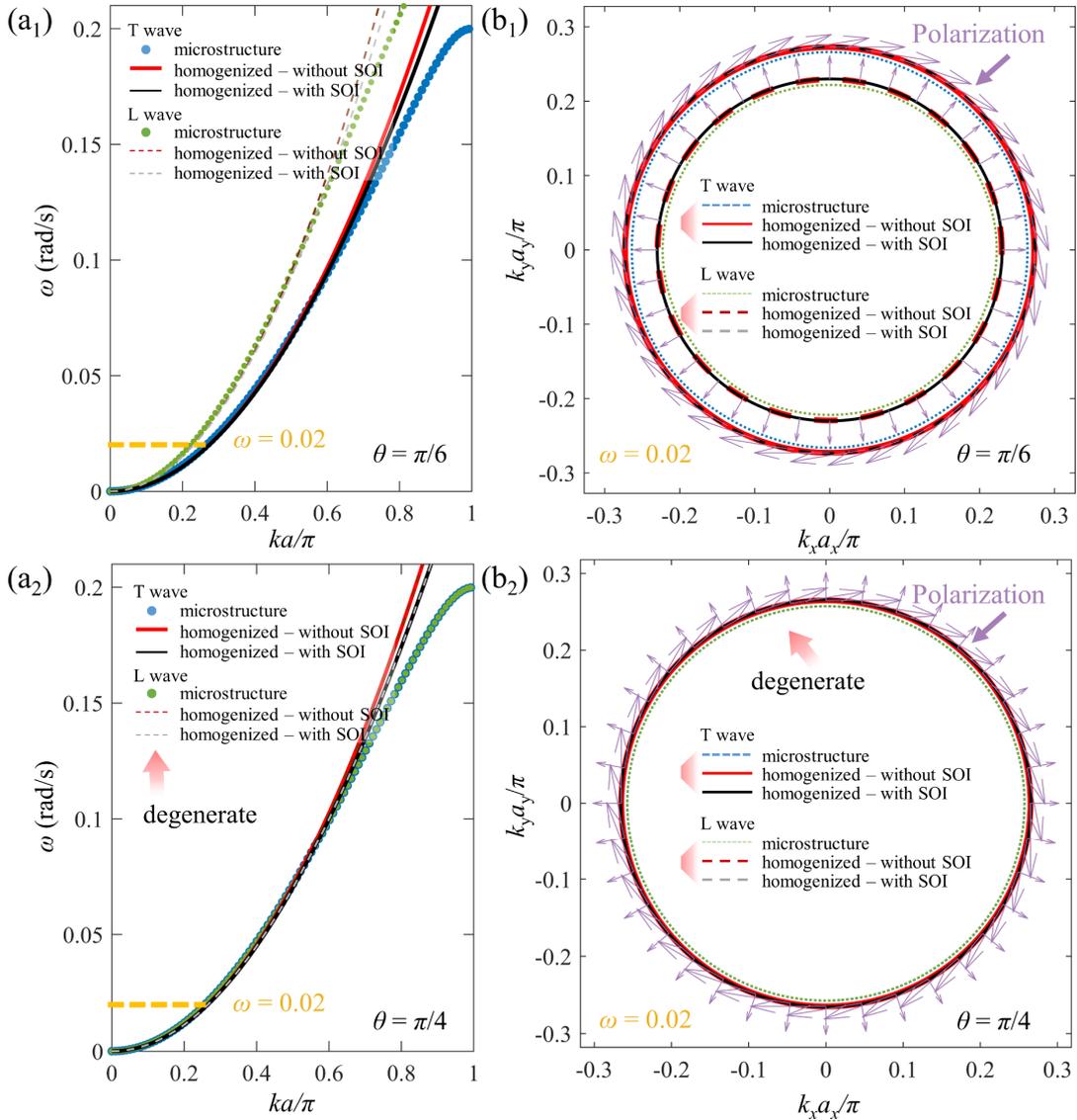

Fig. 6 Bulk wave properties in second gradient materials. ($a_1$) Band structure obtained from the microstructure model (blue and green dots) and the continuum model without second-order inertia term (solid red line and dashed red line) and the continuum model with the second-order inertia term (solid black line and dashed black line). The angle parameter is $\theta = \pi/6$. The generalized "Lamé" parameters, $\kappa = -0.0383$ and $\eta = 0.0683$, for the second-order inertia term are obtained by fitting the dispersion relation of the continuum model to the microstructure calculations.

($b_1$) Equi-frequency curves with $\omega = 0.02$ and the displacement polarization. The displacement of the microstructure model is calculated as the averaged displacement of all 6 explicit nodes in a unit cell. ($a_2$), ($b_2$) Same as ($a_1$), ($b_2$) but for $\theta = \pi/4$. The generalized "Lamé" parameters are $\kappa = -0.0771$ and $\eta = 0.0771$. We remark that the longitudinal waves and the transverse waves are degenerate.

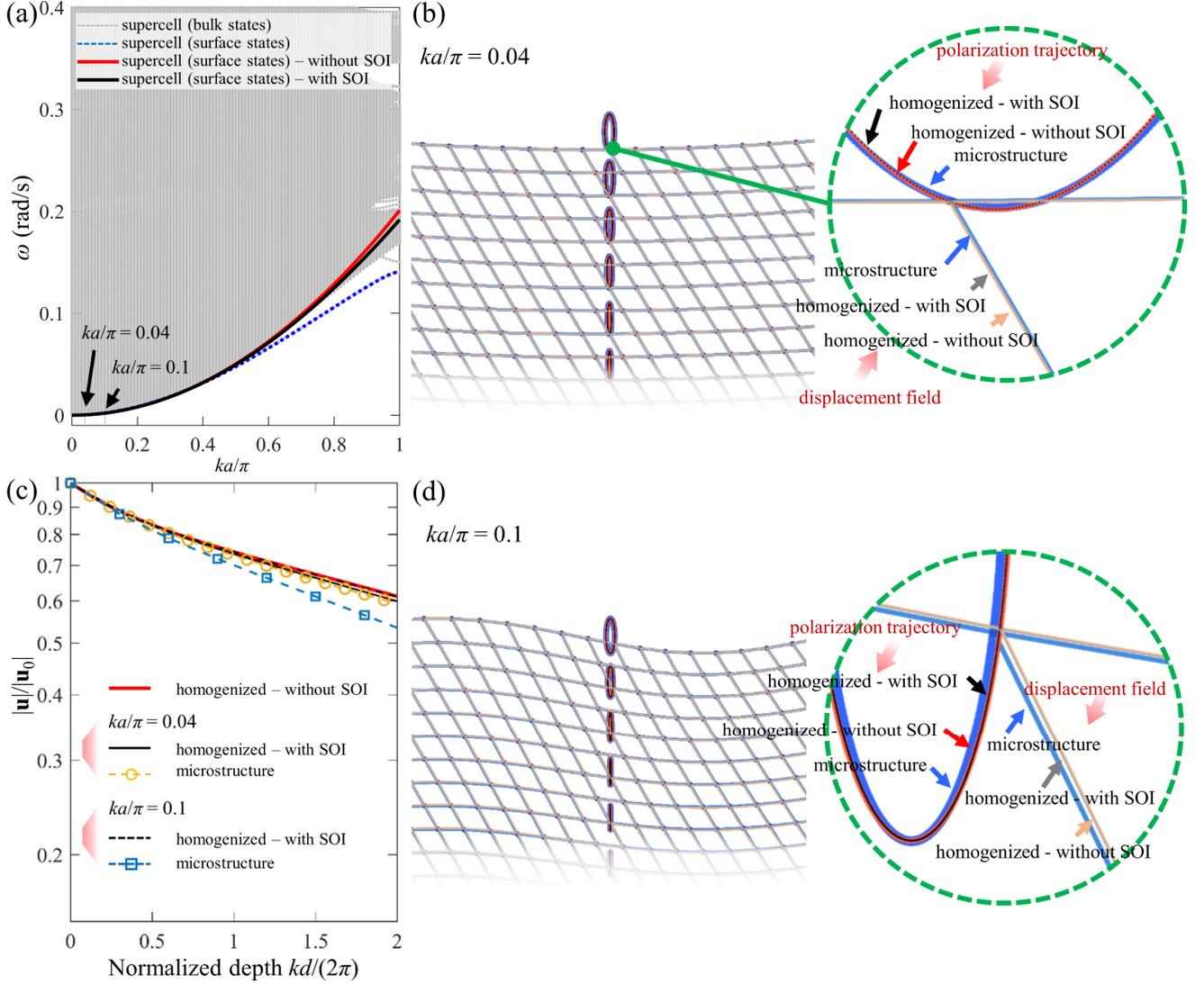

Fig. 7 Verification of one elliptically polarized surface waves in a pure isotropic second gradient material with effective constants $g_2 = g_3 = 0$ and $g_1/g_4 = 0.5$, corresponding to $\theta = \pi/6$ for the microstructure model. (a) Band structure for a strip geometry with 200 unit cells along the $\mathbf{a}_1$ direction (dashed gray lines for bulk states, dashed blue line for surface state), together with surface state results from the continuum model (red line for without second-order inertia, short as SOI, black line for with SOI). (b) Illustration of displacement fields for surface states with $ka/\pi = 0.04$. The blue line, gray line and orange line are for the microstructure model, and the continuum model without SOI, and the continuum model with SOI, respectively. Zoomed-in view plots the displacement trajectories of particles for the microstructure model (blue line), the continuum without SOI (red line), and the continuum model with SOI (black line), respectively. (c) Normalized displacement amplitude $|\mathbf{u}|/|\mathbf{u}_0|$ versus the normalized depth $kd/(2\pi)$ for different $ka$. (d) Same as (b) but for $ka/\pi = 0.1$.

We remark that the physics near free surfaces of architectural materials are generally challenge to describe by using a continuum strain-gradient model (Cakoni et al., 2019; Cornaggia and Guzina, 2020), especially for frequencies away from zero. In literatures, additional boundary layers or boundary correction terms are considered to capture the intricate surface effects or interface effects (Cornaggia and Guzina, 2020; Fergoug et al., 2022; Fliss, 2019). In our study, we focus solely on the long-wavelength limit or the low frequency range. The classical boundary conditions for strain-gradient elasticity derived by Mindlin (Mindlin, 1964) provide a quantitatively good description, as seen from the comparison in Fig. 7.

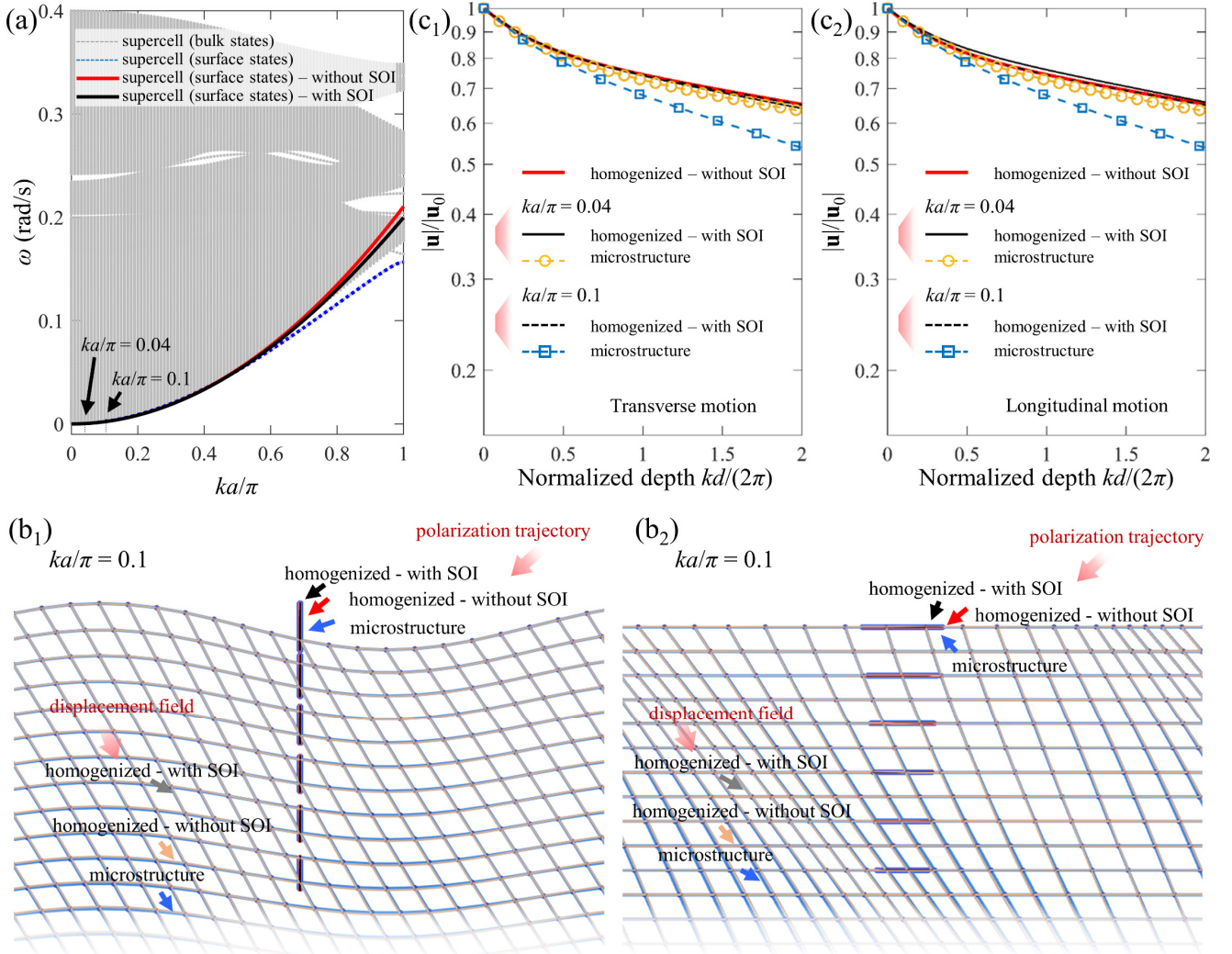

Fig. 8 Verification of two linearly polarized surface waves in a pure isotropic second gradient material with effective constants $g_1 = g_2 = 0$ and $g_3/g_4 = 1$, corresponding to $\theta = \pi/4$ for the microstructure model. (a) Band structure for a strip geometry with 200 unit cells along the $\mathbf{a}_1$ direction. Here, the blue line, the red line, and the black line are each doubly degenerate, corresponding to two degenerate surface states. ($b_1$) ($b_2$) Illustration of the displacement fields for the two surface states with $ka/\pi = 0.1$. ($c_1$) ($c_2$) Normalized displacement amplitude $|\mathbf{u}|/|\mathbf{u}_0|$ versus the normalized depth $kd/(2\pi)$ for different $ka$.

Secondly, we consider the case of $\theta = \pi/4$. The resulting homogenized parameters $g_1 = g_2 = 0$, $g_3 = g_4 = 0.005\sqrt{3}$ correspond to the second example in Section 3.3. Therefore, this isotropic pantographic structure could have two linearly polarized Rayleigh waves, as suggested in the form of Eq. (59). To verify our analysis, we compare in Fig. 8 the dispersion relations and the surface states obtained from the microstructure model and the continuum strain-gradient model. We indeed obtain two degenerate surface modes (see the degenerate blue, red and black line in Fig. 8(a)). The two surface modes have linearly polarized displacement field, in consistent with our previous theory analysis. Again, we obtain quantitatively nice agreement between the continuum strain-gradient model and the microstructure model. Generally, the agreement between continuum model and the microstructure calculations is better for larger angle $\theta$ (see Fig. 4(d)). The larger discrepancy for smaller $\theta$ may due to slight anisotropy of the lattice, since the unit cell is $D_6$-invariant rather than $O(2)$-invariant (Auffray et al., 2015).

## 5. Summary

We have examined Rayleigh waves in 2D extremal materials from both continuum and discrete perspectives. Second gradient high order continuum and its corresponding microstructural model are also proposed to manifest the influence of fine microstructure in these extremal materials. By directly solving the constraint equations, it is demonstrated that Rayleigh waves cannot propagate in 2D extremal elastic materials when the principal material axis is parallel to the free planar surface. However, when the microstructural effect is pronounced, we showed that Rayleigh waves can propagate at finite frequencies. This behavior can be captured by using strain-gradient elasticity. The characteristics of Rayleigh waves in isotropic second gradient elasticity medium have been analyzed in detail. It is found that the polarization of the surface Rayleigh wave in isotropic pure second gradient elasticity can be linearly or elliptically polarized and the Rayleigh waves may not be unique depending on the material parameters. To validate our findings based on strain-gradient continuum theory, we proposed a metamaterial model based on pantographic microstructures. The microstructure calculations agree quite well with the continuum model. This work provides the first systematic study on Rayleigh waves in extremal elastic materials and pure second gradient materials as well, and offers a new perspective to control Rayleigh waves in extremal elastic metamaterials. The analysis here can be generalized for three-dimensional extreme elastic materials, where more possibilities of extreme elastic materials exists, like pentamode and qudramode materials. Other interesting properties might be obtained.

## Acknowledgements

We acknowledge the support from National Natural Science Foundation of China (Grants No. 11632003, No. 11972083, No. 11991030).

## Appendix A

A 3D model of the 1D pantographic unit cell is shown in Fig. A1(a), consisting of two layers, which corresponds to Fig. 5(a). In Fig. A1(a), the rigid rods are linked by ideal joints, and rigid rods with the same color are connected by torsion springs. The structure shown in Fig. A1(b) corresponds to the unit cell of the substructure Group 1 in Fig. 5(b). Finally, Fig. A1(c) shows the toy model of the 2D isotropic pure second gradient elasticity microstructure, which corresponds to the unit cell in Fig. 5(c).

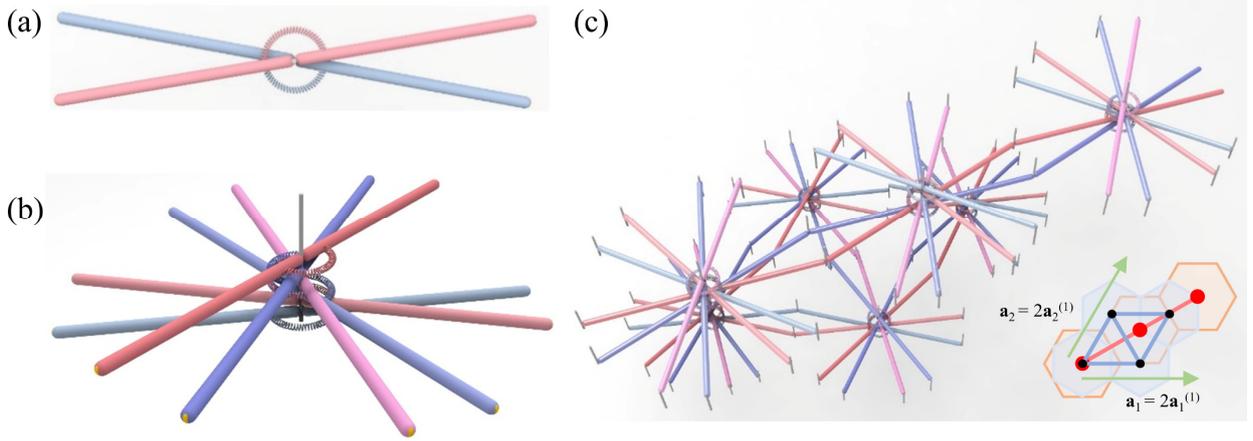

Fig. A1. 2D pure second gradient model with pantographic structures. (a) 1D pantographic truss unit cell toy model corresponding to Fig. 5(a). (b) and (c) are toy model corresponding to the unit cell of the substructure group 1 and the composite unit cell in Fig. 5(c), respectively.

## Appendix B

Here, we give the expression of the sub-stiffness matrices corresponding to torsion spring elements. An 1D pantographic strip in the $(x_1, x_2)$-plane is shown in Fig. B1(a), with $\vartheta$ denoting its angle with respect to the $x_1$-axis. The torsion springs can be simplified sketched as Fig. B1(b).

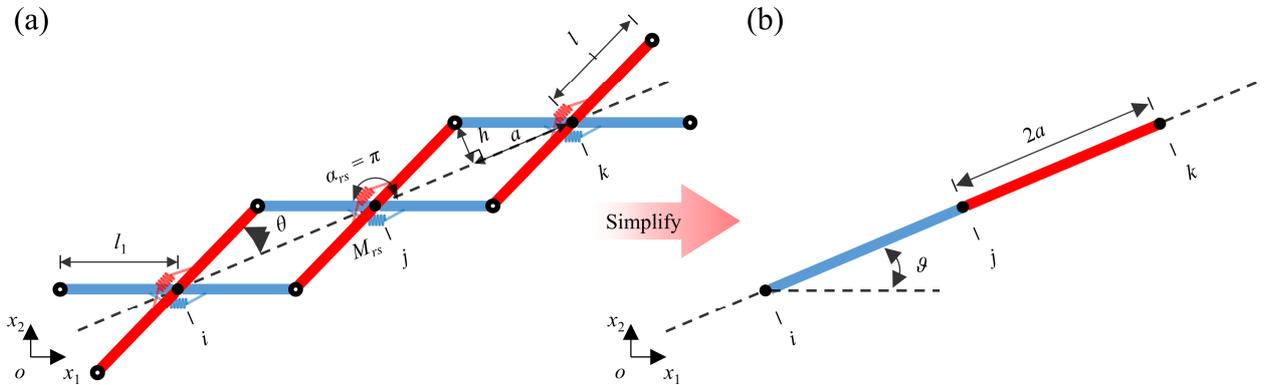

Fig. B1. (a) The 1D composite pantographic strip. (b) The simplified sketch for torsion springs.

When $\vartheta = 0$, the energy of an element in the 1D composite pantographic strip can be expressed as

$$E^{(ijk)} = \frac{1}{2}\left[\kappa_h \left(u_1^{(i)} - 2u_1^{(j)} + u_1^{(k)}\right)^2 + \kappa_a \left(u_2^{(i)} - 2u_2^{(j)} + u_2^{(k)}\right)^2\right], \tag{69}$$

where $\kappa_h = M_{rs}/(2h^2)$ and $\kappa_a = M_{rs}/(2a^2)$ are the effective stiffness of the torsion springs. According to Eq. (69), it can be seen that the energy generated by the torsion spring is contributed by three nodes. The tension at each node can be expressed as

$$\mathbf{f}^{(i)} = \frac{\partial E^{(ijk)}}{\partial \mathbf{u}^{(i)}}, \quad \mathbf{f}^{(j)} = \frac{\partial E^{(ijk)}}{\partial \mathbf{u}^{(j)}}, \quad \mathbf{f}^{(k)} = \frac{\partial E^{(ijk)}}{\partial \mathbf{u}^{(k)}}. \tag{70}$$

Firstly, we consider the stiffness matrix of the torsion springs with an arbitrary $\vartheta$. According to Eq. (70), the stiffness matrix can be naturally expressed as

$$[\kappa]^{(ijk)} = \begin{bmatrix} \dfrac{\partial^2 E^{(ijk)}}{\partial \mathbf{u}^{(i)} \partial \mathbf{u}^{(i)}} & \dfrac{\partial^2 E^{(ijk)}}{\partial \mathbf{u}^{(i)} \partial \mathbf{u}^{(j)}} & \dfrac{\partial^2 E^{(ijk)}}{\partial \mathbf{u}^{(i)} \partial \mathbf{u}^{(k)}} \\ & \dfrac{\partial^2 E^{(ijk)}}{\partial \mathbf{u}^{(j)} \partial \mathbf{u}^{(j)}} & \dfrac{\partial^2 E^{(ijk)}}{\partial \mathbf{u}^{(j)} \partial \mathbf{u}^{(k)}} \\ \text{sym} & & \dfrac{\partial^2 E^{(ijk)}}{\partial \mathbf{u}^{(k)} \partial \mathbf{u}^{(k)}} \end{bmatrix}, \tag{71}$$

where

$$\begin{aligned} \frac{\partial^2 E^{(ijk)}}{\partial \mathbf{u}^{(i)} \partial \mathbf{u}^{(i)}} &= \frac{\partial^2 E^{(ijk)}}{\partial \mathbf{u}^{(i)} \partial \mathbf{u}^{(k)}} = \frac{\partial^2 E^{(ijk)}}{\partial \mathbf{u}^{(k)} \partial \mathbf{u}^{(k)}} = R^T G R, \\ \frac{\partial^2 E^{(ijk)}}{\partial \mathbf{u}^{(i)} \partial \mathbf{u}^{(j)}} &= \frac{\partial^2 E^{(ijk)}}{\partial \mathbf{u}^{(j)} \partial \mathbf{u}^{(k)}} = -2 R^T G R, \\ \frac{\partial^2 E^{(ijk)}}{\partial \mathbf{u}^{(j)} \partial \mathbf{u}^{(j)}} &= 4 R^T G R, \\ R &= \begin{bmatrix} \cos\vartheta & \sin\vartheta \\ -\sin\vartheta & \cos\vartheta \end{bmatrix}, \quad G = \begin{bmatrix} \kappa_h & 0 \\ 0 & \kappa_a \end{bmatrix}. \end{aligned} \tag{72}$$

Then, the global stiffness matrix of a unit cell (or supercell) can be obtained by assembling the sub-stiffness matrices of all elements:

$$[\kappa] = \sum_{i,j,k} [\kappa]^{(ijk)}. \tag{73}$$

Note that the summation symbol in Eq. (73) doesn't mean the algebraic summation but represents the assembly procedure in finite element analysis. Subsequently, particles polarization and dispersion properties for linear wave propagating in an infinite or semi-infinite lattice can be studied by using Bloch theory.